\documentstyle[10pt,emulateapj]{article}
\def\deg{$^{\circ}\,$}
\def\solm{M$_{\odot}\,$}
\def\soll{L$_{\odot}\,$}

\newenvironment{inlinefigure}{
\def\@captype{figure}
\noindent\begin{minipage}{0.999\linewidth}\begin{center}}
{\end{center}\end{minipage}\smallskip}
\submitted{To Appear in Astrophysical Journal Supplements (July 2003)}
\begin{document}

\title{The Relationship Between Stellar Light Distributions of Galaxies and their Formation Histories}

\author{Christopher J. Conselice$^{1}$}

\affil{Henry Robinson Laboratory for Astrophysics, California Institute of Technology, MS 105-24, Pasadena CA 91125}

\altaffiltext{1}{NSF Astronomy \& Astrophysics Postdoctoral Fellow}

\begin{abstract}

A major problem in extragalactic astronomy is the inability to
distinguish in a robust, physical, and model independent way how galaxy 
populations are related to each other and to their formation histories. A 
similar, but distinct, and also long standing question is whether
the structural appearances of galaxies, as seen through their 
stellar light distributions, contain enough physical information to offer 
this classification.  We argue through the use of 240 images of nearby 
galaxies that three model independent parameters measured on a single galaxy 
image reveal its major ongoing and past formation modes, and can be used as 
a robust classification system.  These parameters quantitatively measure: the 
concentration ($C$), asymmetry ($A$) and clumpiness ($S$) of a galaxy's 
stellar light distribution. When combined into a three dimensional `CAS' 
volume all major classes of galaxies in various phases of evolution are 
cleanly distinguished.  We argue that these three parameters correlate with 
important modes of galaxy evolution: star formation and major merging 
activity. This is argued through the strong correlation of H$\alpha$ 
equivalent width and broad band colors with the clumpiness parameter, 
the uniquely large asymmetries of 66 galaxies undergoing mergers, and the 
correlation of bulge to total light ratios, and stellar masses, with the 
concentration index.  As an obvious goal is to use this system at high 
redshifts to trace evolution, we demonstrate that these 
parameters can be measured, within a reasonable and quantifiable uncertainty, 
with available data out to $z \sim 3$ using the Hubble Space Telescope 
GOODS ACS and Hubble Deep Field images.

\end{abstract}


\section{Introduction}

The most basic process in any observational science is the classification
of objects under study into a taxonomy system.  
Advances in scientific fields such as biology and chemistry have historically
depended upon the correct interpretation and ordering of objects 
according to similar, or dissimilar, physical processes or characteristics.  
In 20th century astronomy, the classification of stars via an interpretation
through color-magnitude diagrams led to the correct picture, and new ideas,
for how stars form and evolve.  This has further led to an 
increase in our understanding of stellar systems, most notably
star clusters and galaxies.   A primary concern in astronomy
today is determining how galaxies have formed, evolved, and are evolving. The 
limited progress in our understanding of this is due in no
small part to our inability to identify distinct galaxy populations over
a range of redshifts based on their fundamental features.  If the correct
distinctions can be made, the salient effects of galaxy evolution can be
sorted out and studied, and galaxy formation scenarios should rapidly 
follow.

It is also often believed that galaxy classification systems
are in many ways not very useful for determining evolution.  This 
is the direct result of the most popular classification systems having 
non-physical and often purely descriptive classification criteria based on 
optical morphologies
that do not uniquely identify, or distinguish, galaxies in different
modes of evolution (Appendix A).  This classification crisis is in part the 
result of the high fraction of distant galaxies that look peculiar in deep
Hubble Space Telescope (HST)  images (e.g., Glazebrook et al.
1995; Abraham et al. 1996) and which cannot be placed onto the
Hubble sequence. 
An ideal galaxy classification system is one that classifies objects
on the basis of their most important physical features, which 
the most commonly used morphological systems do not (see Appendix A).

So-called physical morphologies are not new, and attempts to construct a 
meaningful system for galaxies started with the work of Morgan (1958, 1959) 
who attempted to correlate ``the forms of certain galaxies and their stellar 
content as estimated from composite spectra'' from Morgan \& 
Mayall (1957).  Later attempts include e.g., van den Bergh's (1960) study
of the correlation between spiral arm structure and intrinsic galaxy 
brightness and
Morgan and Osterbrock's (1969) classification of galaxies in terms of
dominate stellar populations.
Modern studies have attempted to classify ellipticals by their internal
structures 
(Kormendy \& Bender 1996) and through using interacting and star formation
properties (e.g., Conselice 1997; Conselice et al.
2000a (hereafter CBJ00); Conselice et al. 2000b; Bershady et al. 2000).    
Others have used quantitative features for classification, but without trying 
to understand the physical basis of the measured quantities (e.g., Abraham
et al. 1996; Naim, Ratnatunga \& Griffiths 1997).  However, none
of these systems provide a solid useful frame work for classifying galaxies in
a manner tied to underlying physical processes or properties. 

We argue two main points in this paper.  First, we show that galaxies in 
different stages of evolution can be
distinguished using only three computationally computed
structural indexes.  These three parameters are: the concentration of stellar
light ($C$), its asymmetric distribution ($A$), and its clumpiness ($S$).  All
galaxies fall in various locations of a volume created using these three
structural indices (hereafter CAS volume).  Second, we argue that the values 
of the CAS parameters directly measure the major 
current and past modes of galaxy formation and evolution.  We discuss
evidence that the concentration index traces the past evolutionary 
history of galaxies and potentially correlates with the fraction of stars 
produced 
through gas accreted from the intergalactic medium, as opposed to stars put 
into place through a gravitational infall of gas and pre-existing stars.  
We further argue that the asymmetry
and clumpiness parameters reveal active evolution resulting from
recent major mergers and star formation, respectively, while
the concentration index is the ratio of the integral of these processes.  
The major conclusion from this study is that the structural appearance of 
nearby resolved galaxies, at only a single observed optical 
wavelength, is suitable for understanding their  current 
and past evolutionary history.   The
CAS system can therefore be used to uniquely and automatically identify 
galaxies
of all major morphological types out to $z\sim 3$, including those on the 
Hubble sequence.

This paper is organized as follows:  \S 2 is a discussion of galaxy
structure and its potential usefulness and limitations for uncovering
evolutionary processes in galaxies.  
\S 3 describes the 240 galaxies used in this paper to determine
the usefulness of the CAS system and \S 4 is a description 
of the CAS parameters while \S 5
demonstrates how these parameters correlate with fundamental 
galaxy properties.  \S 6 describes how the CAS
parameters can be used, independent of any physical meaning, to classify
and separate all galaxies into distinct types and argues that we can
use the CAS system at high redshift, and \S 7 is a summary.

\section{The Use of Galaxy Structure: Why a New System is Needed}

When examining the spatial structure of a galaxy in optical, ultraviolet or
infrared light, its appearance is dominated by stars, but effects from
ionized gas and dust play a role, as does projection.  In addition to 
this we have the issue of projection, that is, we view galaxies as two 
dimensional `pictures' when in fact stars are distributed in a galaxy in three 
dimensional space.
When trying to understand a galaxy from its stellar light distributions, and
what it means physically, all of the above effects must be taken into account.

The apparent distribution of stellar light in nearby
bright galaxies quickly led to the Hubble classification system (Hubble 1926) 
which has remained, until recently, the dominant method for understanding 
galaxy populations.  In the last few years however it has been demonstrated 
that classifying
galaxies only using the Hubble sequence has its limitations (see Appendix A).
While nothing a priori should have suggested this - the fundamental
properties of a galaxy are not uniquely revealed through Hubble type 
classification
criteria, i.e., spiral arms and bars.   As such, a new classification system 
which is automated, robust at all redshifts, and which correlates with the 
major physical processes occurring in galaxies is now desperately needed, 
especially
with the advent of the Advanced Camera for Surveys (ACS) and in the near
future WFPC3, both Hubble Space Telescope imaging instruments.
We briefly consider the various options for constructing such as physical
morphology below.

\subsection{Sizes and Shapes}

The most basic structural property of a galaxy is its size, and its two
dimensional shape. The
correlation between the sizes and masses of galaxies suggests that to
first order the size of a galaxy is an important clue for understanding
origins.  Sizes however depend upon knowledge of the distances to 
galaxies, something that is often not know.
 
The importance of the three dimensional shape of a galaxy in terms of its
formation and evolutionary history is not yet entirely clear. 
Statistically, we know that ellipticals and disks must have
different shapes, with disks on average flatter (Lambas, Maddox \& 
Loveday 1992).  Inversion techniques of the
distribution of apparent axial ratios also show that elliptical galaxies are
neither prolate or oblate, but triaxial in shape (Lambas et al. 1992;
Ryden 1996), consistent with a lack of significant rotation 
(Bertola \& Capaccioli 1975; Franx, Illingworth \& de Zeeuw 1991).   

In any case, the detailed three dimensional shape of early type systems may not
be critically important for understanding their stellar content 
and evolution.  This is especially the case for elliptical galaxies 
which have little dust
(White, Keel \& Conselice 2000) and we obtain a clean `view' of its stellar
populations projected onto two dimensions.  The three dimensional shape of a 
spiral galaxy is often inferred from images to be rotationally 
supported disks, an idea confirmed by their
large rotational components (Sofue \& Rubin 2001). 
Despite the fact that some disks are optically thick with
high internal extinctions (White et al. 2000) these galaxies have patchy dust 
distributions, and therefore we are not likely missing a significant amount
of light when galaxies are projected onto two dimensions.  
Later in this paper,
and in CBJ00, we discuss quantitatively how the inclination of spiral galaxies
effects measured morphological properties.
We find that although projection effects are important, their
effect on structure and morphology be can dealt with, and corrected for, in a 
quantitative way (CBJ00, \S 4.3).

\subsection{Light Decompositions}

Understanding the morphology of a galaxy through the decomposition of
its light profile is perhaps the most popular method of quantifying
galaxy light distributions (e.g., Peng et al. 2002;
Simard et al. 2002).
These decompositions are done by fitting a model to surface brightness
profiles, either one or two dimensional.  Typically these models are based 
on various
fitting forms, such as the de Vaucouleurs, exponential or S\'{e}rsic profiles, 
and sometimes also including fitting central point sources (Peng et al. 2002).

These light decompositions are attractive due to their ability to `fit' and
quantify light in `bulge' and `disk' components which are thought to define 
the Hubble 
sequence (Appendix A), although these methods have several problems, both
philosophical and technical.
The most important issue is that bulge and disk decompositions are only 
useful for quantifying the
structures of galaxies with well defined and a priori assumed morphologies, 
namely galaxies with roughly smooth and symmetric light profiles.  There are 
also issues involving the nature of the fitting process which assumes
the form of the light profile and therefore any morphological information is
model dependent.  Galaxies at high
redshifts have peculiar structures that cannot be fit by these
assumed profiles. Since these models are
azimuthally smoothed averages, they do not account for effects caused
by star formation or mergers which can distort a galaxy's light distribution.

It is has also not yet been demonstrated that bulge + disk (B/D) decompositions
are unique descriptions of galaxy light profiles.  Observations 
indicate (MacArthur, Courteau
\& Holtzman 2003) that normal disk galaxies come in several types such that
the relationship between bulge and disk light may not be as
simple as these model fitting techniques assume.  It is also
not clear that bulges are well fit by the de Vaucouleur profile, with 
significant evidence that they are not (Andredakis \& Sanders 1994; 
MacArthur et al. 2003).  If one does not assume a de Vaucouleur or
exponential form, but
use the more general S\'{e}rsic profile, the minimization is over at least 
a dozen parameters
and can result in many different possible solutions.  This is well
known, and B/D decompositions are likely only useful to some degree
for bright nearby galaxies (e.g., MacArthur et al. 2003).  The S\'{e}rsic 
index is also probably not always the best description of a galaxy's light 
profile.  We investigate this
further by modeling the light profiles of galaxies on the Hubble
sequence and compute B/D ratios, and use these to determine how these
relate to our model independent parameters.  

What is needed is a galaxy 
classification methodology that does not rely on any assumptions about galaxy
light distributions and correlates with the major physical processes
occurring in galaxies.  This paper argues that such a system can be 
constructed based on measuring how stars are distributed in galaxies through
the CAS system.

\section{Data and Analysis of Nearby Galaxies}

To develop our CAS physical morphology we study optical digital CCD
images of nearby galaxies in various phases of evolution taken from many
different inhomogeneous data sets.  These include normal galaxies (ellipticals
and spirals), starburst galaxies, dwarf irregulars, dwarf ellipticals
and galaxies undergoing merging.  Each of these separate data sets is
briefly described below.

\subsection{Frei Sample - Nearby Normal Galaxies}

The Frei et al. (1996; hereafter Frei) sample is a collection of 113 nearby
normal galaxies of all classical Hubble types from ellipticals to
irregulars whose quantitative morphological properties are listed in 
Table~1 (see also CBJ00 and Bershady et al. 2000 for
additional physical data on these systems).  Two telescopes were used to 
acquire these
images; the Lowell 1.1m and the Palomar 1.5m.  For the Lowell sample
we use the R-band images and we use the g-band image from
the Palomar sample in our morphological analysis.  We
refrain from a detailed description of the morphological properties of
these galaxies, as this has been done in detail elsewhere (e.g., Frei
et al. 1996; CBJ00; Bershady et al. 2000).  It is sufficient to state
that the Frei sample is a good representation of nearby large
bright galaxies, but is deficient in low-luminosity systems (\S 3.3)
and galaxies in unusual modes of evolution, such as massive star formation
events (\S 3.2) and galaxy mergers (\S 3.4).  

\subsection{Star-Forming Galaxies}

As a significant fraction of galaxies at high redshifts
show morphological and physical evidence for undergoing higher
than average star formation as compared to galaxies in the nearby universe, we
examine the morphological properties of five nearby starburst galaxies
that are generally regarded as the best analogs of high redshift
star-forming systems (e.g., Conselice et al. 2000b,c). These five
galaxies have been discussed in detail in Conselice et al. (2000c).
As a significant fraction of star formation in the high-$z$ universe
is thought to occur in galaxies that are brightest in the infrared, we
also investigate a sample of ultra luminous infrared galaxies that 
are proposed to be the nearby analogs to high-$z$ sub-mm sources (e.g., Blain 
et al. 1999) (see \S 3.4).

\begin{inlinefigure}
\begin{center}
\resizebox{\textwidth}{!}{\includegraphics{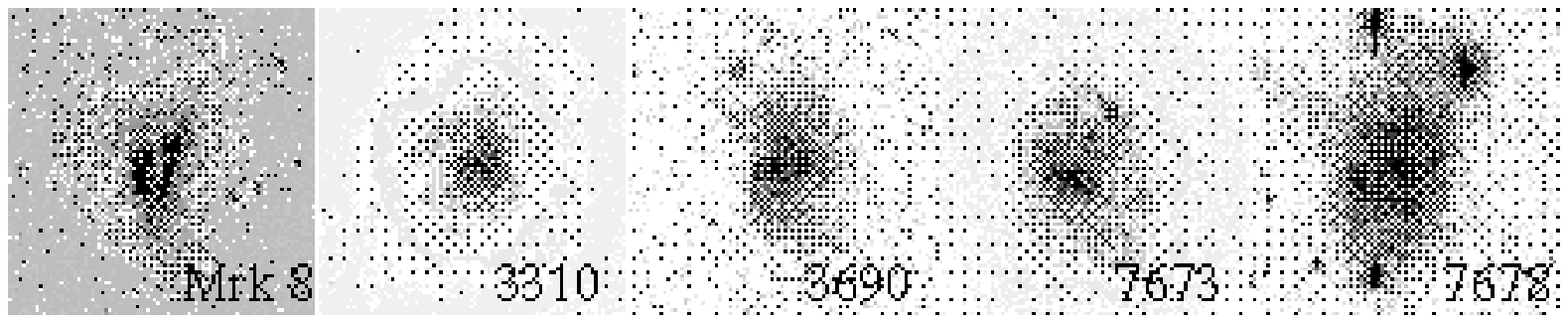}}
\end{center}
\figcaption{Images of the Starburst sample whose properties are listed in Table~2.}
\end{inlinefigure}

Images of these UV bright starbursts are shown in Figure~1 and were all 
observed with the WIYN 3.5 m telescope
in 1998 with the S2kB CCD detector (see Conselice et al. 2000b,c for 
details of the observations).
We use Harris R-band images of each galaxy in
our morphological analyses, the results of which are listed 
in Table 2.  Before performing the structural parameter measurements
on these starbursts, all contaminating features such as foreground stars, 
background galaxies, cosmic rays, and light gradients were removed (Conselice
et al. 2000b). 

\subsection{Dwarf Elliptical and Irregular Galaxies}

Low-mass, or dwarf, galaxies are the most common galaxy type
in the nearby universe, and are particularly abundant in galaxy clusters (e.g.,
Ferguson \& Bingelli 1994; Conselice et al. 2001,2002,2003).  
Any classification system that claims to account for all galaxies
should accommodate these low-mass systems, something the
Hubble sequence does not do (see Appendix A).  As such we include
a sample of 19 dwarf elliptical galaxy images selected from the
sample of Conselice et al. (2003) (Table~3) and 37 dwarf irregulars
from the sample of van Zee (2000) (Table~4).  Like the starburst
sample, these dE galaxies were imaged using the WIYN telescope with the S2kB
detector in the Harris R-band.  The observational details for these galaxies 
is described in Conselice et al. (2003a).  

The dwarf irregular sample are B-band images taken with the 0.9m telescope
at the Kitt Peak National Observatory whose reduction and sample
properties are described in van Zee (2000).  These galaxies were selected
by van Zee (2000) to have M$_{\rm B} > -18$, a late-type morphology, and 
specific HI criteria (see van Zee 2000).  We do not use all the B-band
images from this sample, as some are near bright stars, making a careful
analysis of morphologies difficult.  The 37 galaxies from this sample
that we use are listed in Table~4.  We apply a slight morphological
k-correction, based on the Frei galaxy's B and R morphologies, to convert the 
B-band CAS values of these irregulars into pseudo-R band values.  These 
corrections are however very slight ($\delta C = 0.12, \delta A = -0.03, 
\delta S = -0.03$) and do not effect the results in any significant 
way. The B-band CAS values for these dwarf irregulars are listed in Table~4.

\subsection{Luminous IR Galaxy Sample - Mergers}

Luminous and ultraluminous infrared galaxies (LIRGs) and (ULIRGs) are objects 
that emit
most of their light in the infrared (IR) at $\lambda = 2\mu - 100\mu$
(Rieke \& Low 1972) with luminosities L$_{\rm IR}$ $> 10^{11}$ \soll
(Sanders \& Mirabel 1996). LIRGs and ULIRGs are among the most luminous
objects in the nearby universe bolometrically and in particular
at wavelengths from $8 - 100 \mu$m.   The general idea is that
these galaxies are for the most part recent mergers (Sanders
et al. 1988; Borne et al. 2000; Canalizo \& Stockton 2001) with
induced star formation and AGN activity which produces energetic
photons that are absorbed by dust and later thermally re-emitted
in the infrared by heated grains. 
To address some questions concerning these galaxies several groups obtained
HST images of LIRGS and ULIRGs in the F814W (hereafter I) and F555W 
(hereafter V) bands. We use these samples to study the
morphological properties of merging galaxies, which are much more 
common at high redshifts than in the nearby
universe (e.g., Conselice et al. 2003b).

\begin{inlinefigure}
\begin{center}
\resizebox{\textwidth}{!}{\includegraphics{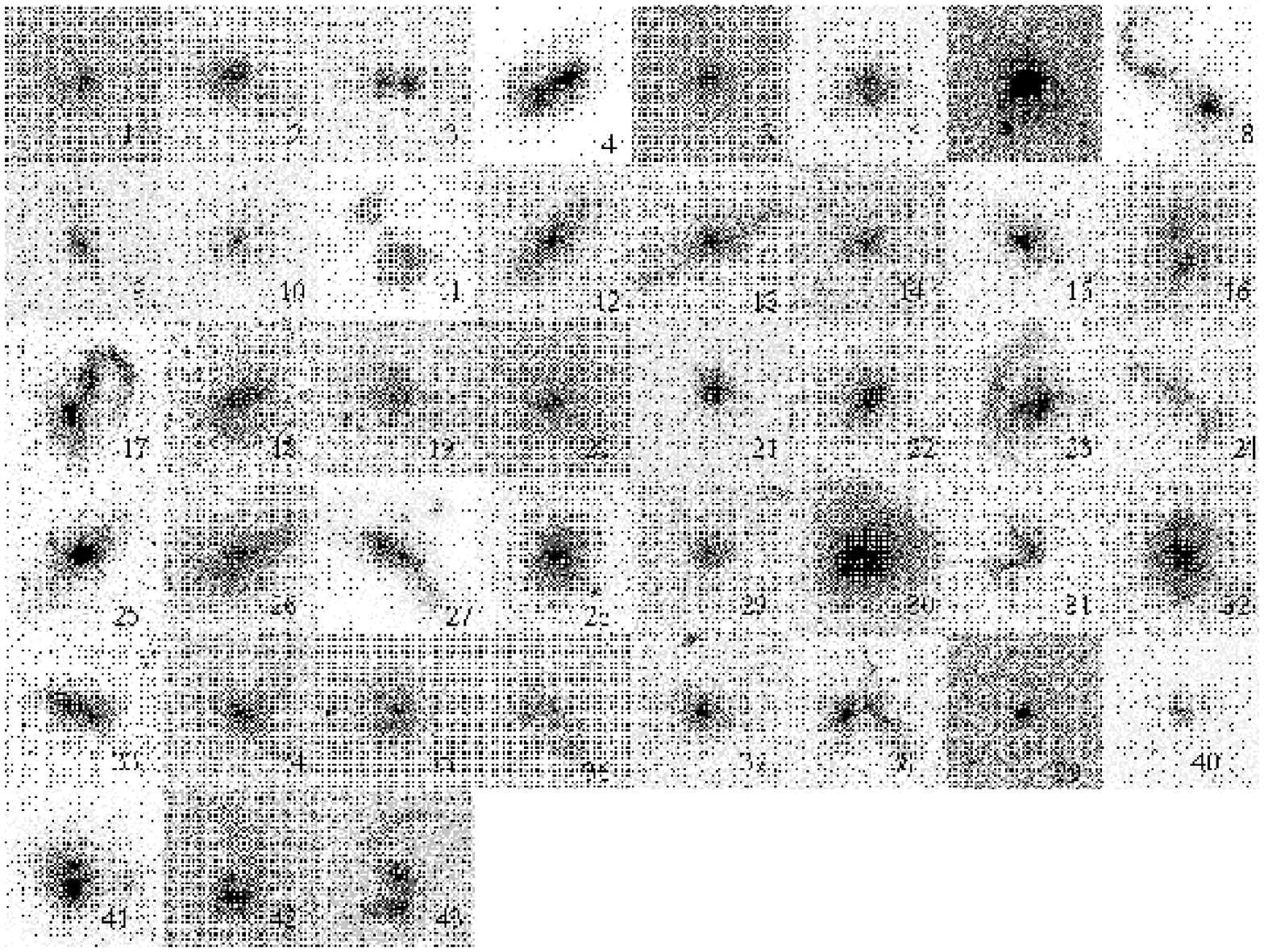}}
\resizebox{\textwidth}{!}{\includegraphics{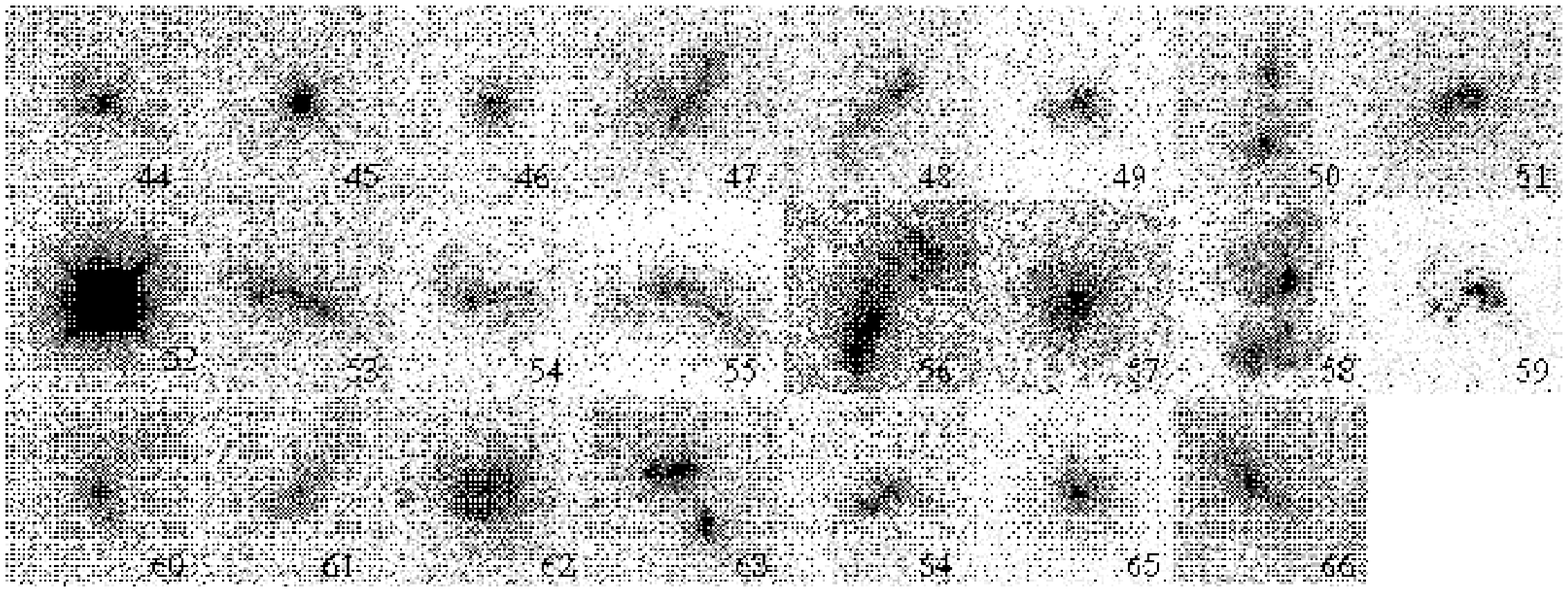}}
\end{center}
\figcaption{Images of the luminous and ultraluminous infrared sample whose properties are listed in Table~5.}
\end{inlinefigure}

Table~5 and Figure~2 list the data and display the images
for the 66 ULIRGs in these two imaging sets (HST IDs 6346 and 6356).  We list
separately, in Table~5, galaxies from these two programs as they use different
filters and exposure times.  The 43 galaxies taken
in the $I$ band were obtained in two separate 400 second WFPC2 exposures and 
two 300 second WFPC2 exposures were acquired for the V-band data. For 
cosmic ray removal the IRAF task CRREJ was used, and the two frames combined.  
We refer to Borne et al. (2000) and Ferrah et al. (2001) for alternative 
descriptions of these galaxies.  As in the dE and
starburst samples, before we apply our morphological techniques we remove any 
contaminating features near the galaxies in question, such as other
smaller background galaxies, foreground stars and any cosmic rays not
removed by the combination of the two separate exposures.  The computed
CAS parameters for these galaxies are listed in Table~5 in the
rest-frame F814W and F606W wavebands.  To compare these values properly with 
the R-band images used for the rest of the sample, we perform a quantitative
morphological k-correction to estimate the rest-frame R band values of
each CAS parameter (see also \S 3.3).  This is done by first computing the 
observed 
wavelength for each galaxy using the filter observed (F814W or F606W) and 
the redshift of the galaxy, which is listed in Table~5.  After this is done, 
the difference in the sampled rest-frame wavelength at F814W or F606W and 
rest-frame R-band at 650 nm is
computed and a morphological k-correction is calculated and applied.  The
value of this correction is found through
an extrapolation based on 
a calibration between the morphological parameters of the Frei sample in
the B and R bands.
\begin{inlinefigure}
\begin{center}
\resizebox{\textwidth}{!}{\includegraphics{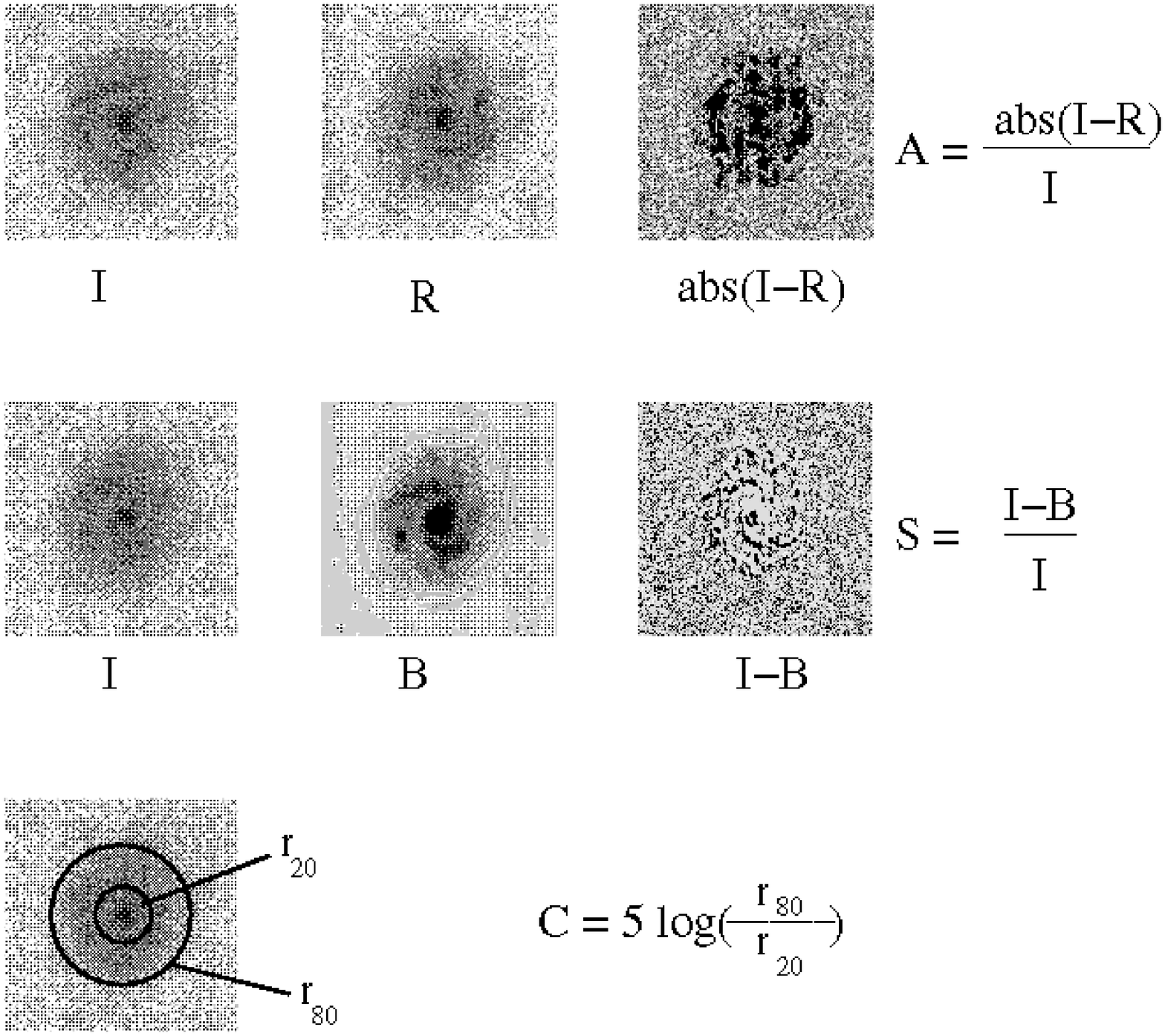}}
\end{center}
\figcaption{Graphical representation of how the three parameters used in this
paper, asymmetry ($A$), clumpiness ($S$) and concentration ($C$) are 
measured.  
For the measurements of $A$ and $S$, I is the original galaxy image, 
R is this image rotated by 180\deg, while B is the image after it has been 
smoothed (blurred) by the factor 0.3$\times$ r$(\eta=0.2)$.
The details of these measurements are not shown here but can be found in
Conselice et al. (2000a) for asymmetry, $A$, Bershady et al. (2000) for
concentration, $C$, and this paper for the clumpiness index, $S$.}
\end{inlinefigure}

\section{The CAS Structural Parameters}

This section describes how the CAS parameters used in this study are
computed.   Figure~3 is
a very basic graphical representation of how the concentration, asymmetry and
clumpiness parameters are measured.   There are possibly better ways
to measure these parameters to strengthen correlations with
physical properties.   In some cases actually measuring the physical
quantities traced by the CAS indices, such as star formation, can supersede
the use of any morphologies.  It is
hoped that this paper will influence others to investigate different
possibilities, such as more automated methodologies of measuring
galaxy structures through e.g., shapelets (e.g., Massey et al. 2003). Later 
in the paper we
discuss what internal physical properties the CAS parameters correlate with,
and construct a three-dimensional structural space and 
show that all major nearby galaxy types can be cleanly distinguished within 
this volume.

\subsection{Concentration of Light}

Light concentration is an often used feature in galaxy
classification studies, and has been used qualitatively previously in
the Hubble system, and in various alternative classification schemes
(e.g., Morgan \& Mayall 1957). Briefly, elliptical galaxies are the most
concentrated systems, and the concentration of stellar light decreases for
later Hubble types (e.g., Bershady et al. 2000). 
Disk galaxies with large rotational velocities, and high angular momenta, 
have lower light concentrations (e.g., Bershady et al.
2000), as do dwarf galaxies (Conselice et al. 2002).   We further
discuss the relationship between concentration and physical properties of
galaxies in \S 5.

Measuring the light concentration quantitatively
can be done by using a single index which has been
described in detail in many papers, using a variety of different techniques
(e.g., Morgan \& Mayall 1957; Okamura, Kodaira \& Watanabe 1984; Kent 1985;  
Doi, Fukugita \& Okamura 1993; Abraham et al. 1994; Bershady et al. 2000; 
Graham, Trujillo \& Caon 2001).  In this paper the same methodology described 
in Kent (1985) and Bershady et al. (2000) is used.     
This concentration index is a number, $C$, defined 
as the ratio of the 80\% to 20\% curve of growth radii (r$_{80}$, r$_{20}$),
within 1.5 times the Petrosian inverted radius at r($\eta = 0.2$),
normalized by a logarithm, $$C = 5 \times {\rm log(r_{80\%}/r_{20\%})}.$$ See 
Bershady et al. (2000) for a full description of how this parameter is
computed.   The asymmetry (\S 4.2) and clumpiness (\S 4.3) 
parameters are also measured within the 1.5 $\times$ r($\eta = 0.2$) 
radius (see CBJ00 for a discussion 
of the benefits of using this particular value).
The quantitative values of $C$ range from roughly 2 to 5 with
most systems having $C > 4$ ellipticals, or spheroidal systems in formation, 
with disk galaxies at values generally between $4<C<3$.  
The lowest concentrated objects are those with low central
surface brightnesses, and low internal velocity dispersions (Graham et al.
2001; Conselice et al. 2002a). 

\subsection{Asymmetry}

The asymmetry index, $A$, has been used previously to quantify the 
morphologies
of galaxies, usually those seen at high redshift (e.g., Shade et al. 1995;
Abraham et al. 1996; Brinchmann et al. 1998; Conselice et al. 2003b).  
However,  there are only a few previous
attempts to calibrate and characterize an asymmetry
index on nearby galaxies, and models of galaxy formation, to understand
its importance for measuring galaxy evolution 
(Conselice 1997; Conselice et al. 2000a,b; Conselice 2000; Conselice
et al. 2001). 

The asymmetry index is defined as the number computed when a galaxy
is rotated 180\deg from its center and then subtracted from its
pre-rotated image, and the summation of the intensities of the absolute value 
residuals of this subtraction are compared with the original galaxy flux.
There are further corrections applied for removing the background, and
well-defined radii and centering algorithms (see CBJ00 for a full detailed 
description). The asymmetry index used in this paper,  as defined in 
CBJ00, is sensitive to any feature that produces asymmetric  
light distributions.  This includes
star formation, galaxy interactions/mergers, and projection
effects such as dust lanes.   Since most galaxies are not edge-on systems,
star formation and galaxy interactions/mergers are likely the dominate 
physical effects (CBJ00).   While no non-mergers have very high asymmetries the
corollary is not true, and galaxies that have undergone
a merger can have modest asymmetry values (see \S 5.2.1).  Asymmetry
also tends to correlate with the $(B-V)$ color of galaxies, an indication that
it is sensitive to some degree to the ages of a galaxy's
stellar populations.

\begin{inlinefigure}
\begin{center}
\resizebox{\textwidth}{!}{\includegraphics{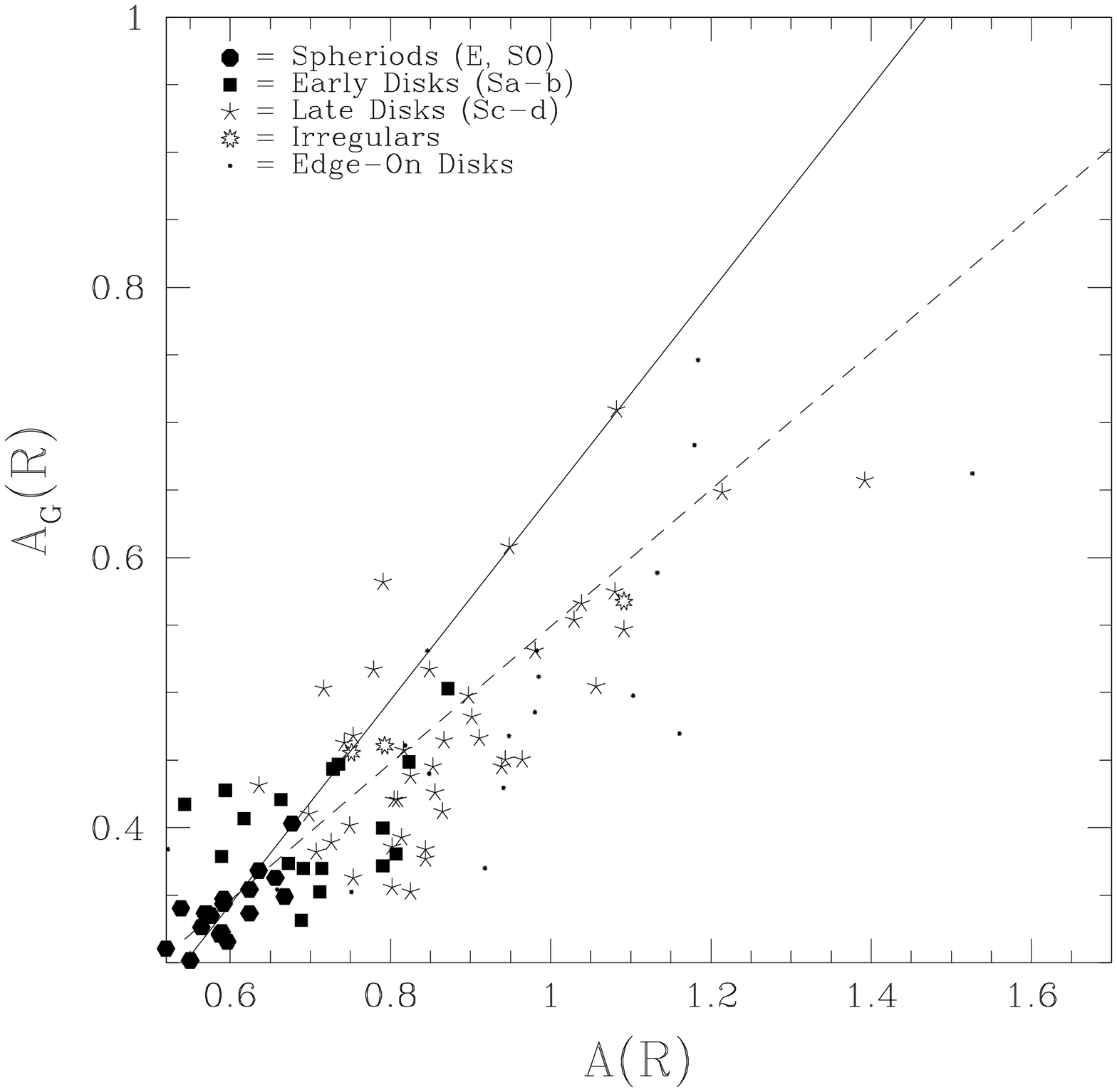}}
\end{center}
\figcaption{The global asymmetry parameter $A_{G}(R)$ plotted as a function
of the asymmetry parameter $A(R)$.  The dashed line is a best fit between
these two parameters, while the solid line shows the relationship
$A_{G}(R) = A(R)$.}
\end{inlinefigure}

It would be ideal to have an asymmetry index sensitive
only to large scale stellar distributions.
To measure this we compute the asymmetries of each 
galaxy image in our sample after it is convolved with a smoothing 
filter of size 1/6 $\times$ r($\eta$ = 0.2), to reduce the image's
effective resolution.  The asymmetry of this smoothed
galaxy is then computed in the normal manner (CBJ00).  We call this
value the global asymmetry, $A_{\rm G}$(R).
This is equivalent to studying the asymmetric components of a galaxy's
low-frequency structure.\footnote{Ideally a 
measurement of the global asymmetry would be computed by
considering all light from a galaxy equally.  That is, all pixels in
a galaxy are given a binary value of 1 or 0 depending on whether or
not light is being emitted from the galaxy at each respective pixel. This is
decided by determining if pixel values are greater than, or less than, 
some level of the sky variation
$\sigma$.  Everything less than n$\sigma$ is set to zero while everything
greater than n$\sigma$ is set to 1 and is presumably a pixel that samples part 
of the galaxy.  Although this method was tried using several $\sigma$ values 
and various radii, the best
turned out to be a 3$\sigma$ division.   The resulting image then contains only
pixels with values 0 or 1.   The global asymmetry
index was then computed using the same techniques presented in
CBJ00.  However, after extensive testing it was found that this procedure 
was not robust, and gave global asymmetry measurements that correlated strongly
with the type of galaxy, with elliptical galaxies
usually having the highest values due to the systematics caused by
their extensive surface brightness envelopes.  Therefore
this methodology was rejected for further use.}

Figure~4 shows the relationship between the global asymmetry, $A_{\rm G}$(R) 
and the total asymmetry $A$(R) for the Frei sample.  There is a very good 
correlation between these two parameters, such that

\begin{equation}
A_{G}(R) = (0.67\pm0.05)\times A(R) + (0.01\pm0.01).
\end{equation}

\noindent Based on this, only $\sim$ 30\% of galactic asymmetries can be 
produced 
by high frequency structures, such as star formation.    Because the 
asymmetry of a galaxy
generally does not decrease strongly with decreased resolution, its
asymmetric structures are likely caused by global distortions.   
If star formation was
a significant factor in producing asymmetries, the 
$A_{G}$ values would decrease significantly from the original $A$ value.  
Since they do not, as was also argued in CBJ00, 
it is likely that star formation is not producing a large asymmetry
signal, and we simply use the pure asymmetry index, $A$(R),
to obtain an estimate of global asymmetries.

\subsection{The High-Spatial Frequency Clumpiness Parameter ($S$)}

The light distribution in a galaxy can be roughly characterized by
three features: its integrated light distribution, the rotational
symmetry of this distribution and the patchiness of this distribution 
(Appendix~B).
The first two are described by the well studied asymmetry and 
concentration indices, 
while the third can be measured with a parameter 
introduced here, which we call the clumpiness ($S$).   
Earlier work on similar techniques includes investigations 
by Isserstedt \& Schindler (1986) and Takamiya (1999). 

There are several ways a galaxy can contain clumpy material 
at high-spatial frequencies.
Ellipticals, for example, usually do not have any 
high-spatial frequency power in their structures.  Most ellipticals are 
`smooth' and 
therefore only contain low-frequency, or a nearly 
DC low-power structure.  However, galaxies undergoing star formation are 
very patchy and contain large amounts of light at high-frequency.  
To quantify this, we define the clumpiness index $(S)$ as the ratio of the 
amount of light contained in high frequency structures to the total amount 
of light in the galaxy.  For ellipticals this ratio should be, and generally
is, near zero. 

After experimentation the best and
most simplistic way to measure the high spatial frequency components in
a galaxy was found.  The computational method is simple:
the original galaxy's effective resolution, or `seeing' is reduced
to create a new image.  This is done by smoothing the galaxy by
a filter of width $\sigma$.  The effect of this is to created an image
whose high-frequency structure has been washed out.  This
can also be thought of as lowering the $\epsilon$ resolution parameter of 
CBJ00  such that one seeing element contains some fraction of a galaxy's
radius.   The original image is then subtracted by this new reduced
resolution image.  The effect of this is to produce a residual map 
that contains only the high frequency components of the galaxy's stellar
light distribution (see Figure~3).  The flux of this residual light
is then summed and divided by the sum of
the original galaxy image flux to obtain a galaxy's clumpiness ($S$) value.  
If the subtraction process results in
few residuals, it implies that the galaxy contains little light at high
frequencies and the $S$ index is near zero.  On the other
hand, if a significant fraction of the light in a galaxy originates
from these high-frequency clumpy components, the $S$ index will be large.  
Computationally $S$ is defined as:

\begin{equation}
S = 10 \times \Sigma_{\rm x,y = 1,1}^{\rm N,N} \frac{(I_{\rm x,y} - I^{\sigma}_{\rm x,y}) - B_{\rm x,y}}{I_{\rm x,y}}
\end{equation}

\noindent where $I_{\rm x,y}$ is the sky subtracted flux values of the galaxy
at position (${\rm x,y}$), $I_{\rm x,y}^{\rm \sigma}$ is the value of the
galaxy's flux at ({\rm x,y}) once it has been reduced in resolution by
a smoothing filter of width $\sigma$, and N is the size of the galaxy
in pixels.  The value $B_{\rm x,y}$ is the 
background pixel values
in an area of the sky which is equal to the galaxy's area.  
Computationally the clumpiness index, $S$, has additional measurement
features.
The inner part of each galaxy is not considered in the
computation of $S$ as these are often unresolved and
contain significant high-frequency power that is unrelated to the
stellar light distribution, but an artifact of finite sampling.  We also 
force any negative
pixels, that is parts of a galaxy that are over subtracted in equation
(2), to zero before computing $S$.    We also use a smoothing filter of size 
$\sigma(I)$ = $1/5 \times 1.5 \times r(\eta = 0.2$) = 0.3 
$\times r(\eta = 0.2)$, although in principle any scale
can be used to measure various frequency components in a galaxy.  The $S$
values for our various galaxy types are listed in Tables 1 - 5.

\section{Physical Interpretation of the CAS Parameters}

\subsection{Light Concentration - Scale and Past History}

In major galaxy formation scenarios 
the concentration of a galaxy's light strongly depends upon its formation 
history, that is how its stars were put into place.
In hierarchical galaxy formation
scenarios elliptical galaxies and spiral bulges form by the mergers of 
pre-existing 
stellar and gaseous systems which lose angular momentum through the ejection 
of tidal tails of material. This results in high stellar light concentrations.
Alternatively, in a monolithic collapse scenario, stars are concentrated
in an elliptical galaxy or in a bulge due to the rapid collapse of gas that 
quickly forms stars in less than 100 Myrs, or from a rapid collapse of 
pre-existing stars.
In these models the resulting spheroids can then acquire a disk from star 
formation produced 
from gas deposited from the intergalactic medium that cools onto the
spheroid (Steinmetz \& Navarro 2002).  After the gas inside each halo 
cools, its angular momentum  
determines the properties of the formed disk 
including its light concentration, which will be lower than systems formed
through a rapid gravitational collapse. 

Does the concentration index really tell us about any fundamental 
galaxy properties, or how a galaxy has formed?    
A galaxy's light concentration has been shown since Okamura, Kodaira \&
Watanabe (1984) to 
correlate with the internal properties of galaxies, notably Hubble
types and surface brightness.  Recent studies have also shown that
light concentration indices correlate with internal scaling properties 
such as velocity dispersion, galaxy size, luminosity and black hole mass 
(Graham et al. 2001).  There are also various definitions of the 
concentration index, such as the central light 
concentration (Graham et al. 2001) 
that correlate better with some of these parameters.  Investigating various
light concentration indices, including which is the best for measuring scale
features is beyond the scope of this present paper, but see Graham et al.
(2001) for a good discussion of these issues.

\subsubsection{Concentration vs. Bulge to Total Light Ratios}

The concentration index we use does however correlate with
several internal galaxy properties.  First, the concentration index scales
with the bulge to total light flux ratios (B/T) for nearby normal galaxies
(Figure 5), although there is a large scatter, particularly
for early type disk galaxies.  We measure these B/T ratios for the
sample of normal Hubble sequence galaxies in the Frei sample using the
GALFIT fitting software
(Peng et al. 2002).  GALFIT is a 2-dimensional bulge + disk $\chi^{2}$ 
minimization program that finds the best de Vaucouleur and exponential
profile fit to the light distribution of a galaxy (Peng et al. 2002). 
Figure~5 shows the relationship between C and B/T whose best fit 
relationship is,

\begin{equation}
{\rm B/T} = (0.4\pm0.03) \times C + (-0.88\pm0.11).
\end{equation}

\noindent The concentration index can thus be used as an indicator
of a galaxy's gross form (i.e., elliptical or spiral), 
within some scatter.  The 
origin of this scatter is not clear as the residuals of the linear fit
in eq. (3) do not correlate with the inclination, asymmetry or any other
morphological feature. 

\begin{inlinefigure}
\begin{center}
\resizebox{\textwidth}{!}{\includegraphics{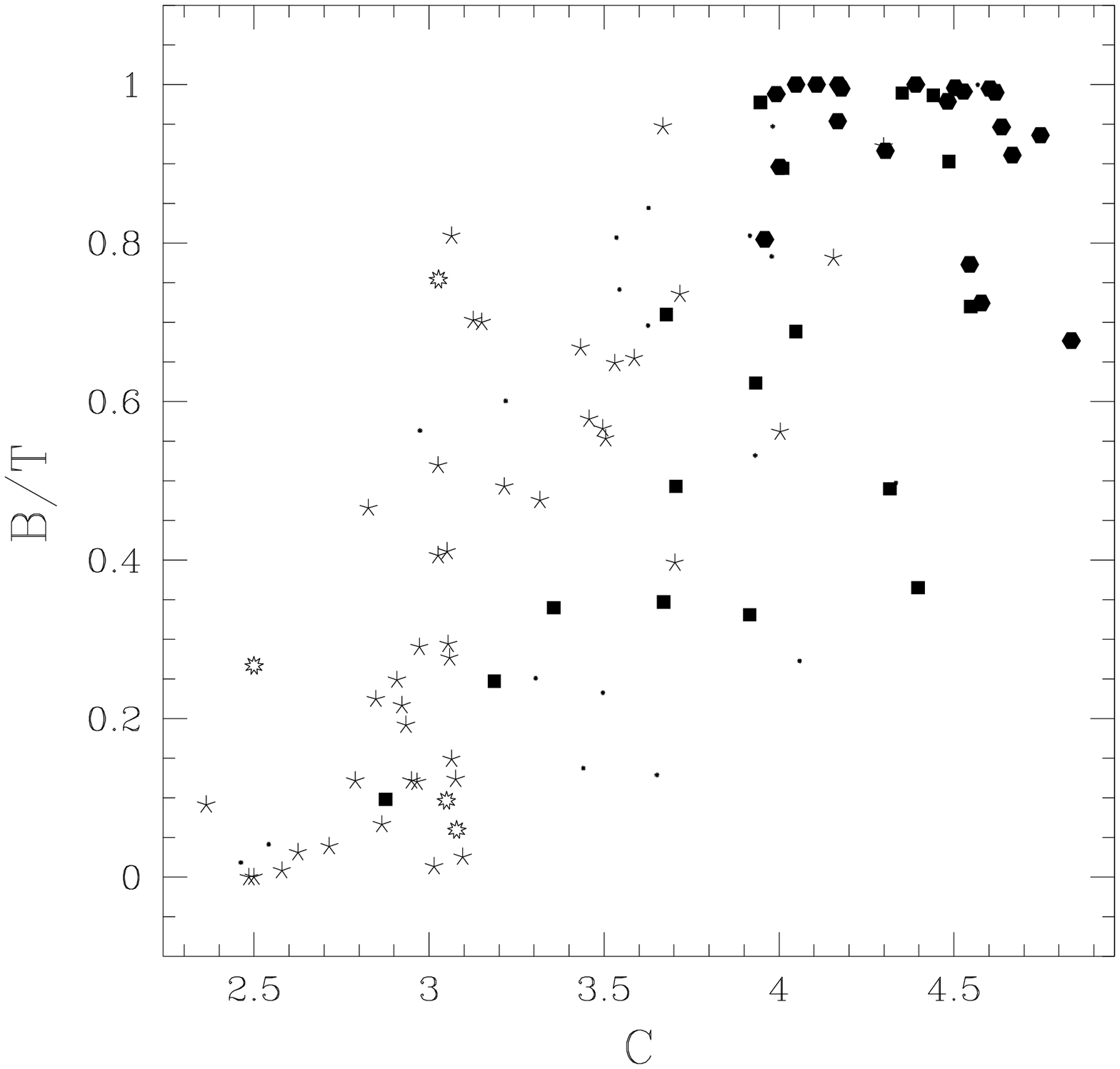}}
\end{center}
\figcaption{Bulge to total light ratios for the Frei sample as a function of 
the total light concentration.  The symbols here have the same meaning
as in Figure~4.}
\end{inlinefigure}

As shown in Graham et al. (2001) the concentration index is also a measure
of the scale of a galaxy, whereby we mean the size, absolute magnitude or
mass, as defined in Bershady et al. (2000).   Graham et al. (2001)
argue that the central concentration of light is a better measure of 
scale than total light concentration, although there is still a correlation 
between the total light concentration and the stellar mass, as given by
the factor M$_{\rm B} - (M/L)_{\rm B}$, where $(M/L)_{\rm B}$ is the stellar 
mass to light
ratio in the B-band estimated from the $(B-V)$ color of each galaxy (Worthey
1994).   
When we compare this value to the concentration index, $C$, we find a general 
correlation,
with an average scatter of one magnitude (Figure~6).  This relationship is:

\begin{equation}
{\rm M_{\rm B}} = (-2.0 \pm 0.3) \times C + (-14.2\pm 1.1).
\end{equation}

\begin{inlinefigure}
\begin{center}
\resizebox{\textwidth}{!}{\includegraphics{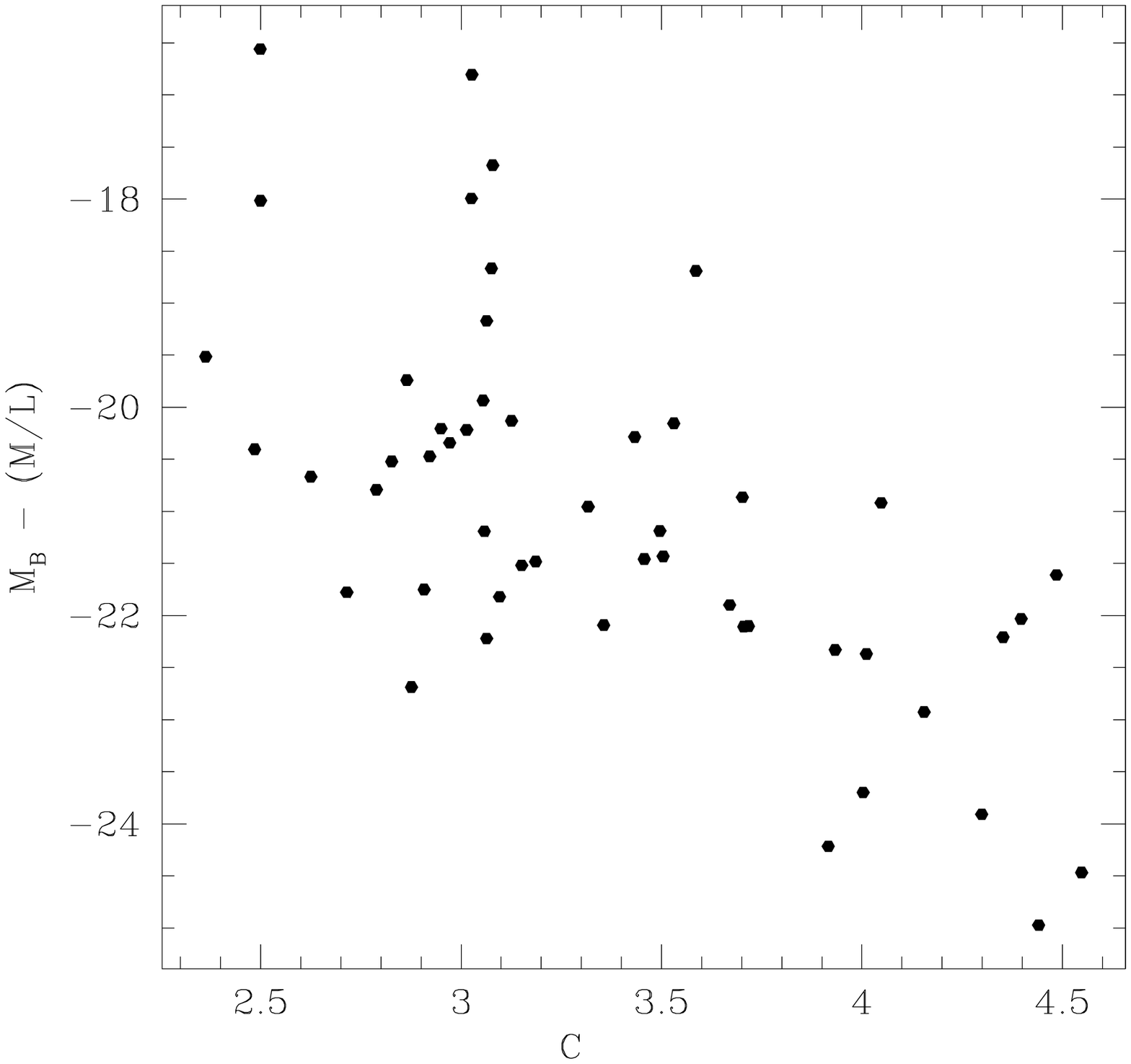}}
\end{center}
\figcaption{The relationship between the concentration index, $C$, and
M$_{\rm B}$ - M/L for the nearby normal galaxies in the Frei sample.}
\end{inlinefigure}

\noindent Based on this, and the results of Graham et al. (2001), it seems 
very 
likely that the scatter shown in Figure~6 can be significantly reduced by 
using either a refined concentration index, or through the use of a careful 
analysis
using a third morphological parameter that effects light concentration, such
as inclination or star formation.
 
\subsection{Asymmetry - Mergers}

Galaxy interaction and merger rates, and the fraction of galaxies involved
in mergers, is predicted in hierarchical models to
increase as a function of redshift (e.g., White \& Frenk 1991; Lacey \& 
Cole 1993; Cole et al. 2000).  Mergers in this scenario build up the
masses and structures of galaxies.  Observations suggest that 
mergers are indeed more common at higher redshifts (Conselice et al. 2003b).

Despite this, the importance of galaxy mergers in the galaxy evolution
process is still debated, although galaxy mergers are clearly 
occurring.  The effects of past merging activity is
easily seen in the local galaxy population.  For example, Fourier analyses of 
nearby galaxy structures reveal that
over 30\% of spirals show evidence for recent accretion events
(Zaritsky \& Rix 1997). 
Our own galaxy and M31 also contain evidence for past merging events, such as
a heated thick disk and stellar streams that are likely the result of 
accretion of satellites (Quinn, Hernquist \& Fullagar 1993; Gilmore,
Wyse \& Norris 2002).  Evidence for past mergers can also be found in 
the outer parts of galaxies where dynamical time
scales are long, and where apparently normal galaxies have evidence for
past merger activity in the form of stellar shells (Schweizer \& Seitzer 1992)
and HI asymmetries and warps (Sancisi 1976; Richter \& Sancisi 1994).
While many galaxies have some clues left over from past merging events, these
can be quite diverse, and require various methods to retrieve, many of which 
are expensive in telescope time, and hard to impossible to
measure at high redshifts.

\begin{inlinefigure}
\begin{center}
\resizebox{\textwidth}{!}{\includegraphics{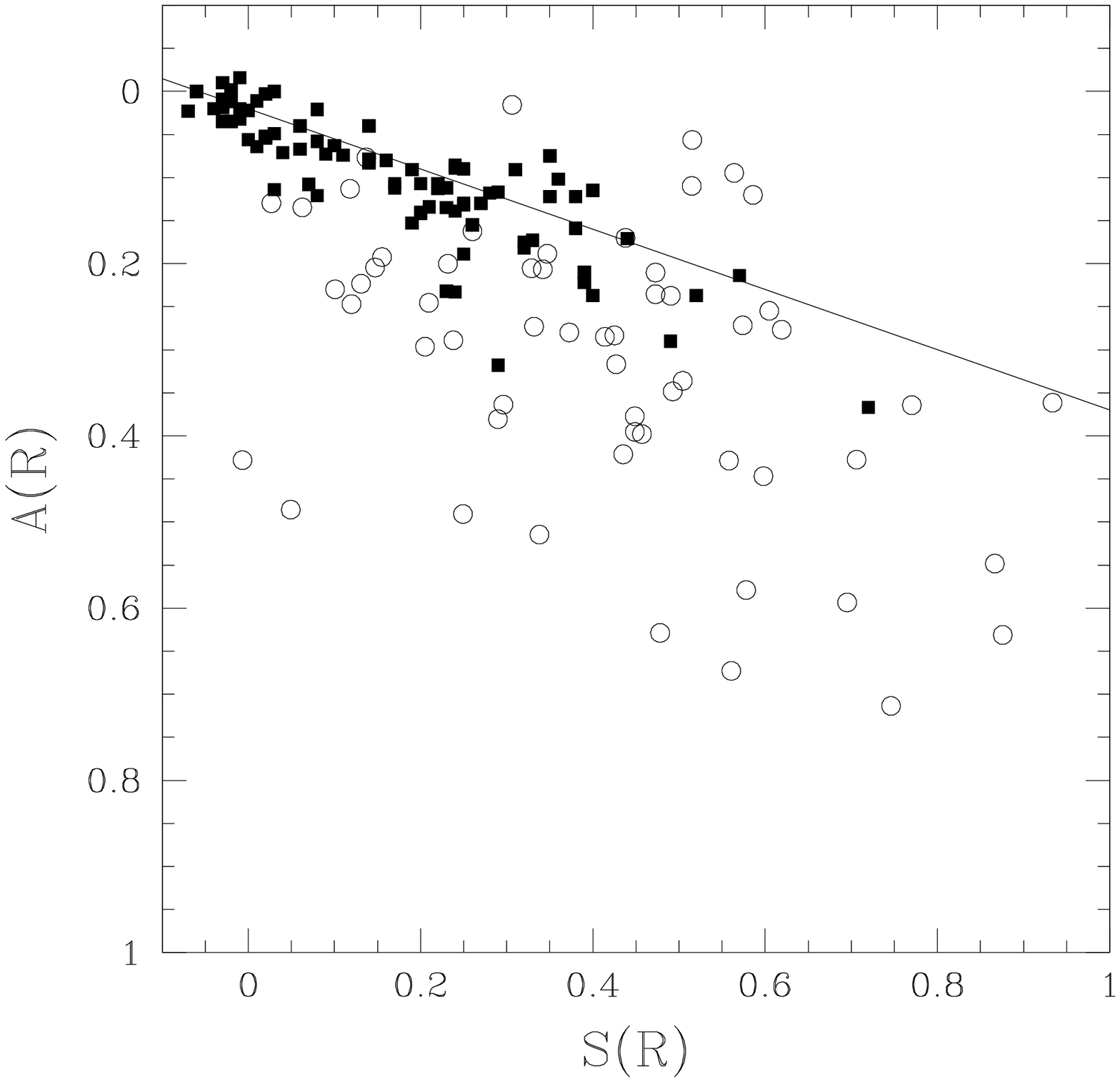}}
\end{center}
\figcaption{The relationship between the asymmetry, $A$, and clumpiness 
parameter, $S$, for the normal Frei galaxies (solid boxes) and the ULIRGs 
(open circles).}
\end{inlinefigure}

It has been previously argued that the asymmetry 
index described in \S 4.2 is a good and simple morphological
indicator of galaxies undergoing galaxy interactions and mergers
(CBJ00; Conselice et al. 2000b). The question remains, however, as to how 
good this structural index 
is at distinguishing between mergers in different stages, and how 
it can be used to determine if a galaxy is undergoing a minor or major 
merger. To answer this, we first examine the asymmetries of the 113 nearby
bright normal Frei galaxies spanning all Hubble morphologies (\S 3.1), none
of which have been involved in recent major mergers.    
As was shown in CBJ00, the asymmetries of the non-edge on Frei galaxies 
correlate with their broad-band $(B-V)$ colors (Conselice 1997; CBJ00).  
The asymmetries of these galaxies also correlate with the clumpiness index, 
$S$ (\S 5.3) (see Figure~7).  The fit between asymmetry and $(B-V)$ color 
for normal Hubble galaxies is,

\begin{equation}
A(R) = (-0.35\pm0.03)\times ({\rm B-V}) + (0.35\pm0.03),
\end{equation}

\noindent while the correlation between the asymmetry and the clumpiness
index, $S$, is given by:

\begin{equation}
A(R) = (0.35\pm0.03)\times S(R) + (0.02\pm0.01). 
\end{equation}

\noindent The relationship between $A$ and $S$ for the Frei and
ULIRG sample is shown in Figure~7, while the relationship between $A$ and
$(B-V)$ is discussed in detail in CBJ00. 
For normal galaxies that are not highly inclined, both of these 
correlations with $A$ have a well defined scatter in asymmetry at various
$(B-V)$ color and clumpiness ($S$) ranges.  Figure~8 shows the difference
between the Frei sample asymmetry values and the fits given in eq. (5) and 
(6), as a function of $(B-V)$ and $S$.  

The average scatter in these differences for both correlations 
is plotted as a function of $S$ and $(B-V)$ as open circles at two 
different bins, high and low $S$ and $(B-V)$.  These 1$\sigma$ 
scatter
values are roughly constant at all colors and clumpiness values, and are 
$\sigma(A) \sim 0.035$ for both fits.

As galaxies that are undergoing mergers appear more distorted in their
stellar and gaseous distributions than a normal
elliptical or spiral, the asymmetry parameter for these galaxies should
intuitively increase and deviate from these fits.  Indeed, this has
been confirmed by studying a small number of systems undergoing mergers
(Conselice et al. 2000b).  Here we investigate this idea
more systematically.

\begin{inlinefigure}
\begin{center}
\resizebox{\textwidth}{!}{\includegraphics{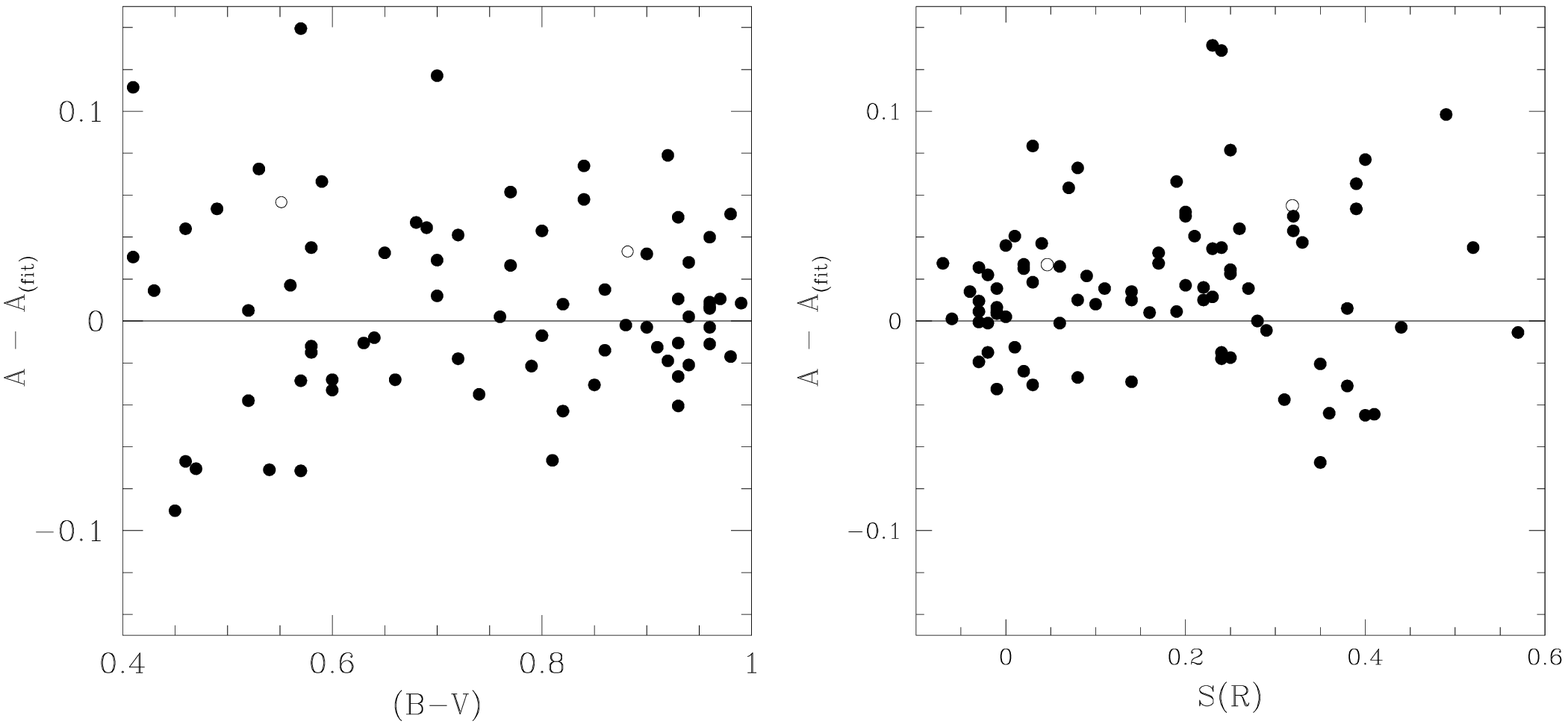}}
\end{center}
\figcaption{Relationship between the asymmetry deviation from the fits in
eq. (5) and (6) for the Frei sample as a function of the $(B-V)$ and
$S(R)$ parameter for each galaxy.  The open circle is the scatter of
this deviation at low and high $S$ and $(B-V)$ values.}
\end{inlinefigure}

A merger will not in general increase the clumpiness $S$ or color
as significantly as it does the asymmetry, as galaxies
that are non-interacting can be dominated by star formation as much
as galaxies that are involved in a merger (e.g., Conselice 2000b; \S 5.3).
This is supported by the starburst/merger observations presented in \S 3
where the $S$ index for star-forming galaxies (starbursts
and ULIRGs) are roughly the same as irregular and starburst galaxies not 
involved in recent mergers (\S 6).  
Furthermore, as no normal galaxies deviate significantly from the
fits given in eq. (5) and (6) we argue that galaxies that deviate 
due to their large asymmetries, are involved in major mergers.   As a limit
we use $(A - A_{\rm predict}) > 3\sigma$ from the $A$ vs. $S$ and 
$A$ vs. $(B-V)$ correlations to identify objects that are 
undergoing major mergers.   We test this idea below by using 
images of the ULIRGs, most of which are generally thought to be involved in 
mergers.

\subsubsection{ULIRGs as Mergers}

How well does the asymmetry index, as used in the CAS system, do in 
identifying 
ULIRGs as mergers?  Figure~9 shows a histogram of the asymmetry values for
the Frei and ULIRG galaxies.  While the Frei galaxies have mostly
low asymmetries, the ULIRGs span a much larger range and include
systems at all $A$ values, including many highly asymmetric ones.   
Thus, the asymmetry index fails to uniquely identify all ULIRGs as
mergers since many have asymmetries as low as the normal Frei
galaxies.  This reveals that the asymmetry index is not
sensitive to all phases of the merging process, or that not all
ULIRGs are mergers.  
Systems that have recently undergone a merger, or those in the beginning
of a merger, do not have high asymmetries, as can be 
shown using N-body models of galaxies involved in the merger process 
(Conselice \& Mihos in prep).

We can test the idea that the asymmetry parameter is measuring only 
the central phase of a major merger by plotting the IR luminosity of
our ULIRG sample with the $\sigma$ deviation from the asymmetry-clumpiness 
correlation.  This comparison
is shown in Figure~10 where we plot the deviation from the $A-S$
relationship (eq. 6) in units of $\sigma$ deviations as a function of the
ratio of the 100$\mu$m
and 60$\mu$m fluxes for our ULIRG sample.  
Systems with the highest asymmetry values are those that have slightly
higher ratios of F$_{100}$/F$_{60}$.  This may demonstrate that galaxies
with the highest relative 100$\mu$m fluxes have the largest asymmetries, 
consistent with the idea that these galaxies have heated dust from
more active AGN and starbursts induced by mergers.

\begin{inlinefigure}
\begin{center}
\resizebox{\textwidth}{!}{\includegraphics{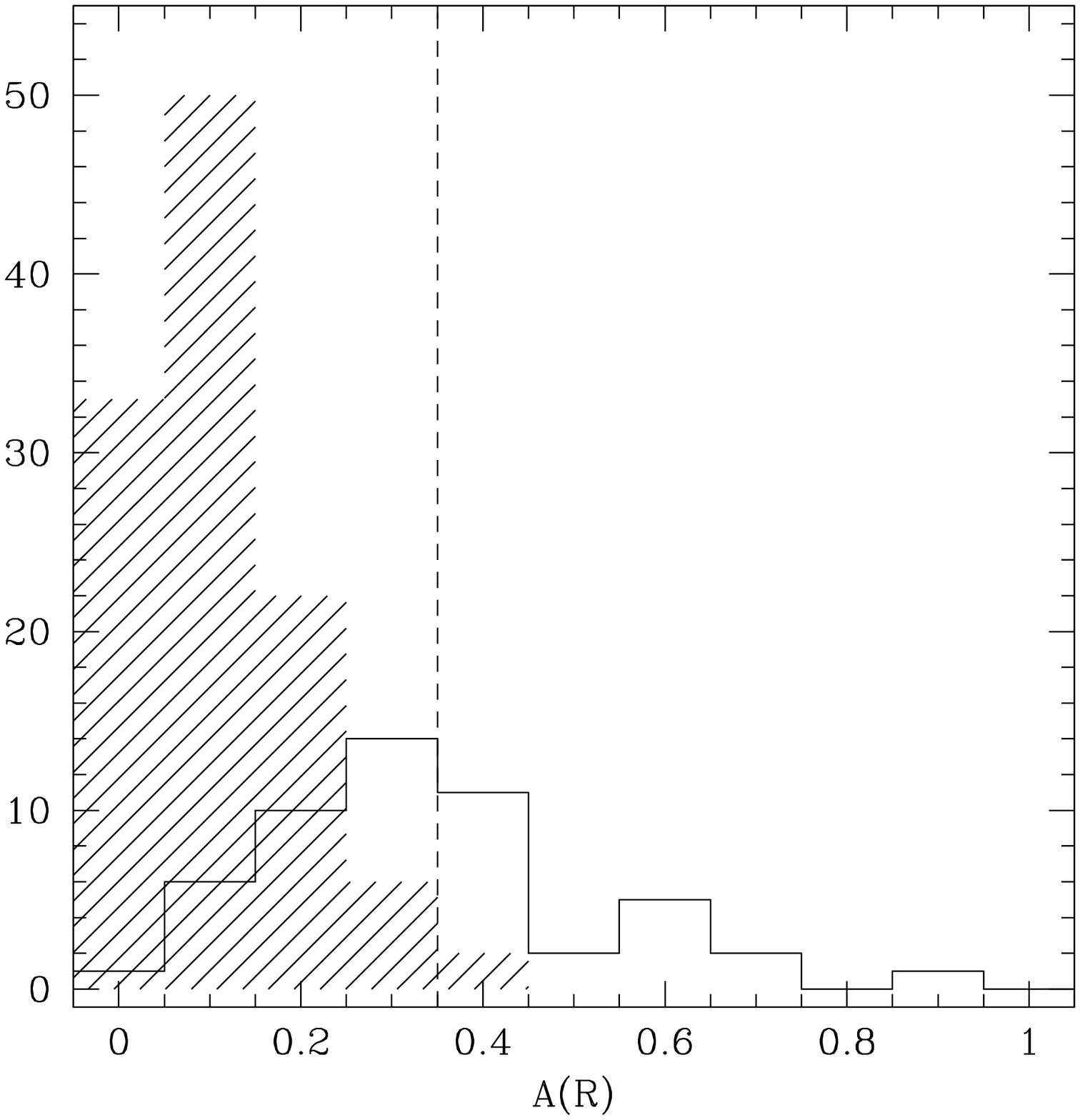}}
\end{center}
\figcaption{Histogram of asymmetries for the Frei galaxies (shaded) and
the ULIRGs (open).   }
\end{inlinefigure}

Figure~11 shows the asymmetry deviation in $\sigma$ units for our ULIRGs, 
and Frei galaxies, from the $A-S$ correlation (Figure~7) as a function of 
clumpiness
($S$).  Most systems that deviate at $> 3\sigma$ 
are the ULIRGs, with only a few irregulars and late-types higher than this 
limit.  However, as mentioned earlier, not all the
ULIRGs have asymmetries high enough such that they would all be
identified as ongoing mergers.  In fact, about 50\% of the ULIRGs have 
asymmetries consistent with being involved in ongoing major
mergers, i.e. $\delta$A $> 3\sigma$. This is consistent with eye-ball
estimates of the fraction of ULIRGs involved in major mergers (Borne et al.
1999).  If the
HST sample of 66 ULIRGs we use is representative of galaxies involved in
various stages of the major merger process, then the total number of galaxies
undergoing a major merger in any sample would be underestimated by a factor of
two by using the asymmetry methodology.  

\subsection{Clumpiness - Star Formation}

There is a morphological dichotomy between galaxies dominated
by recent star formation and those that are not, and this is one of the major
physical processes that defines the Hubble sequence, and most other
galaxy morphology schemes. Traditionally,
ellipticals, both normal and dwarf, have little star formation whereas spirals 
and irregulars often have morphologies dominated
by it, especially when they are observed at ultraviolet, or 
optical wavelengths.  For example, the morphologies of late-type spiral 
galaxies in optical and ultraviolet light are dominated by bright young stars 
(e.g., Windhorst et al. 2002).   The presence,
or absence, of star formation is also a fundamental parameter for understanding
galaxy evolution, simply because galaxies are made of stars, and all stars
were once young and therefore the passive evolution of stellar populations
will have a strong effect on galaxy morphology.
It is therefore desirable to find a simple method based on the
appearances of galaxies that can be used to understand their
star formation properties.  We argue here that the
clumpiness parameter is a powerful method for doing this.

\begin{inlinefigure}
\begin{center}
\resizebox{\textwidth}{!}{\includegraphics{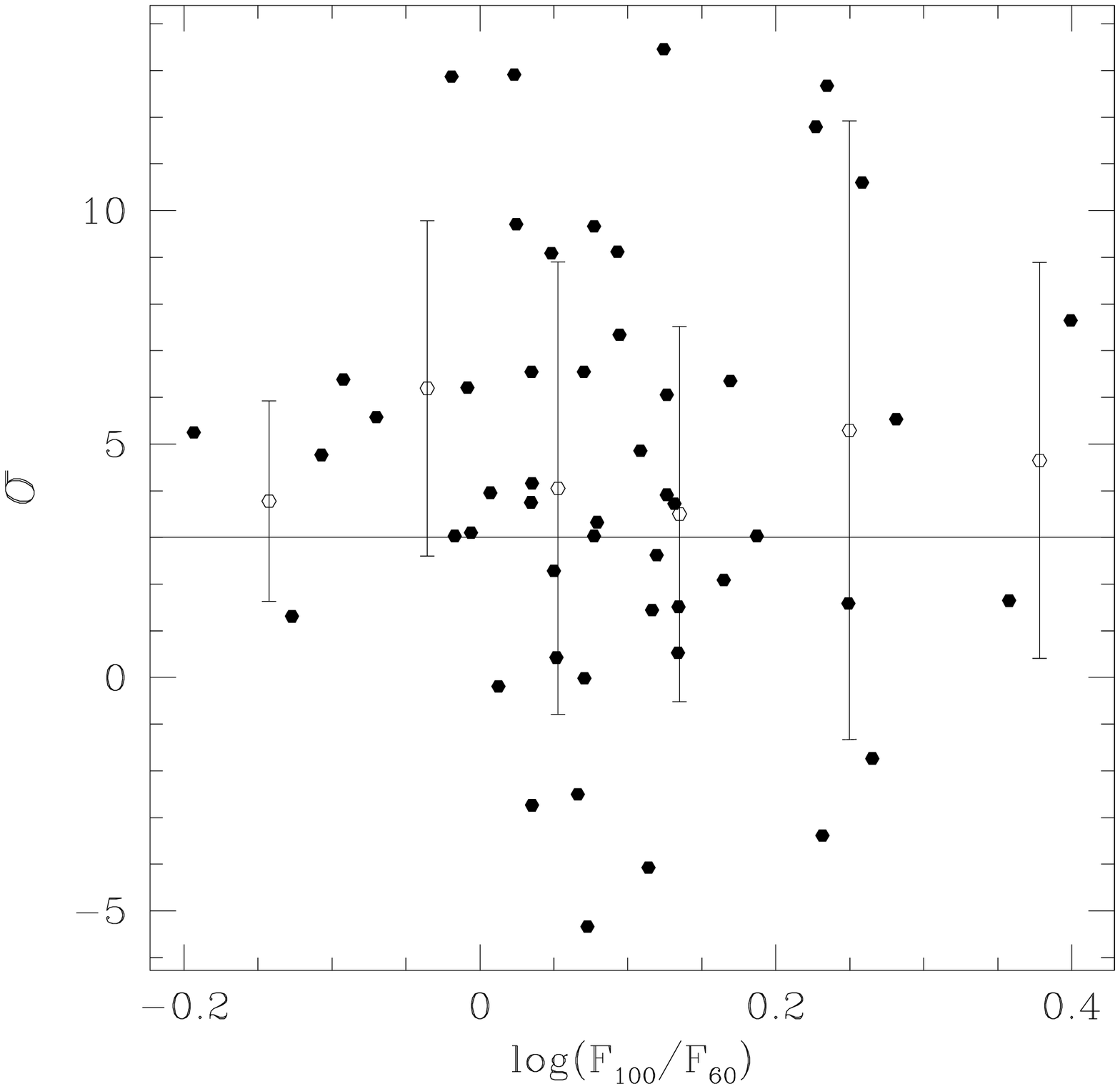}}
\end{center}
\figcaption{The deviation in sigma ($\sigma$) units from the 
asymmetry-clumpiness relationship for the ULIRGs as a function of the ratio 
between the IRAS flux at 100 $\mu$m and 60 $\mu$m.}
\end{inlinefigure}

When viewing images of nearby galaxies at a fixed resolution   
ellipticals are the smoothest type, while spirals and irregulars have
a clumpy appearance due to star clusters and stellar associations that are the 
sites of recent star 
formation.  The clumpiness of a galaxy's light is partially the result of 
this clustering.   As nearly all stars appear to form in 
star clusters which later dissolve (Harris et al. 2001) the idea that the 
clumpiness of a galaxy's light correlates with star formation is a natural 
one. The best way to argue that the clumpiness parameter correlates with the
presence of young stars is to determine how $S$ 
traces measures of star formation, such as H$\alpha$ fluxes or 
integrated colors sensitive to stellar ages.
Figure~12 shows the relationship between the clumpiness ($S$) values of the 
Frei galaxies and their
$(B-V)$ colors taken from de Vaucouleurs et al. (1991) and their 
H$\alpha$ equivalent widths from Kennicutt \& Kent (1983) and Romanishin
(1990).   From this, the clumpiness index ($S$) correlates with both $(B-V)$ 
and H$\alpha$ equivalent width, but both with a large scatter.  

\begin{inlinefigure}
\begin{center}
\resizebox{\textwidth}{!}{\includegraphics{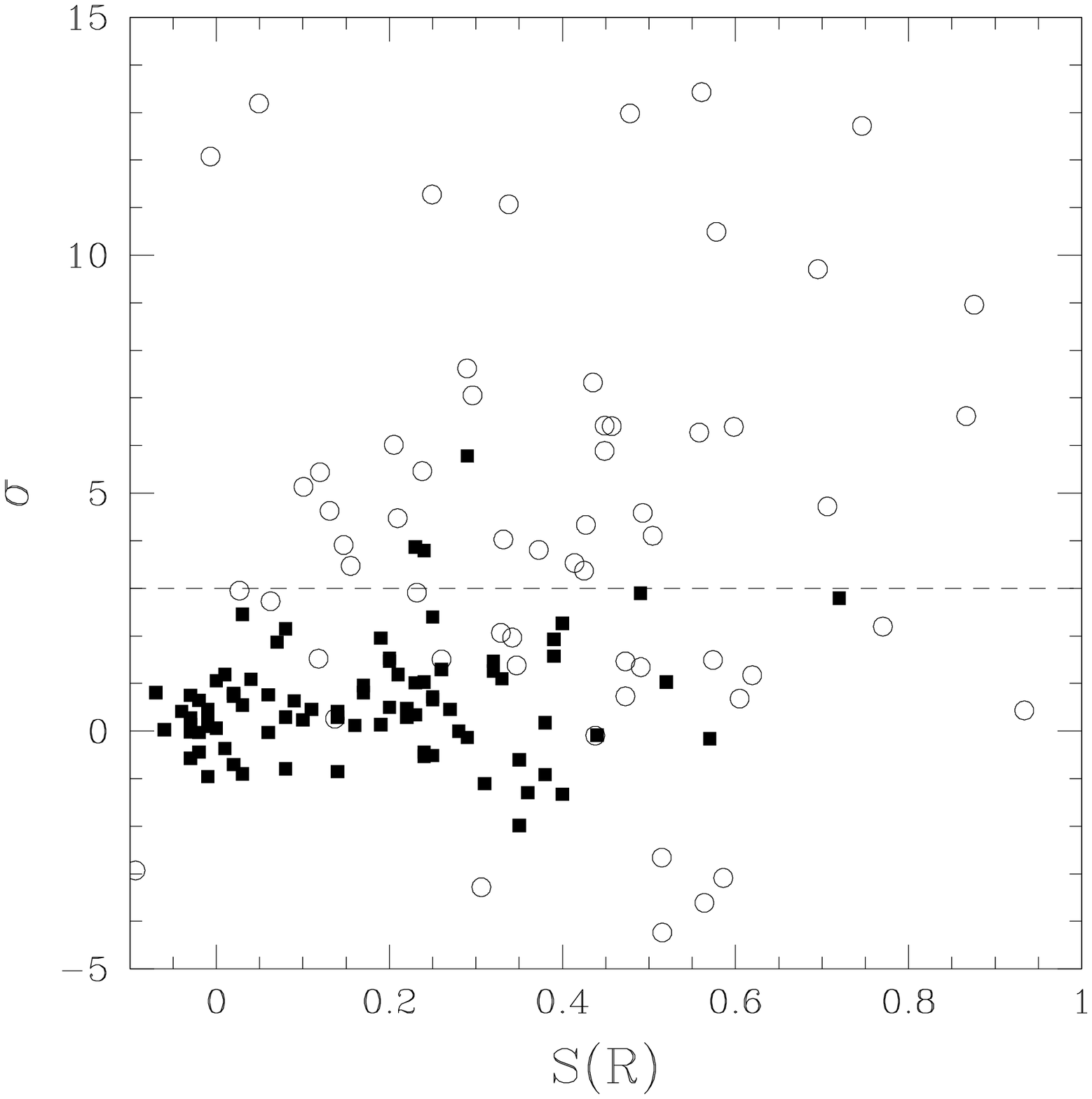}}
\end{center}
\figcaption{The sigma $\sigma$ deviation from the the asymmetry-clumpiness
relationship as a function of the clumpiness value ($S$) for ULIRGs
(open circles) and the Frei galaxies (solid boxes).   }
\end{inlinefigure}

This scatter is largely due to the effects
of different viewing angles and projection effects.  That is, the clumpiness 
parameter ($S$), due to 
its morphological nature, is sensitive to dust lanes and inclination.  
Figure~13 shows the residuals of the fit
between $(B-V)$ and H$\alpha$ equivalent width with $S$, which we call 
$\delta S$, as a function of axis
ratio, $\epsilon$ = log (a/b) as defined in de Vaucouleurs et al. (1991).   
The galaxies 
that deviate the most from this linear fit are those with the highest 
inclinations, or largest axis ratios.
The fit between $\epsilon$ and the scatter in the $S$ fit, $\delta S$, can be 
represented as power laws 

\begin{equation}
\delta S_{(B-V)} = (0.010\pm0.006) \times (1 + \epsilon)^{7.2\pm1.1},   
\end{equation}

\begin{equation}
\delta S_{H_{\alpha}} = (0.004\pm0.004) \times (1 + \epsilon)^{8.2\pm1.0}.
\end{equation}

\noindent After calculating an inclination corrected $S$ parameter by 
subtracting out these terms, the result of which we call $S' = S - \delta S$,
we find that the correlations 
between $S'$ and $(B-V)$ and H$\alpha$ significantly improves (Figure~14).  The
relationship between $S'$ and H$\alpha$ equivalent width and $(B-V)$ colors is
given by the best fit relationships,

\begin{inlinefigure}
\begin{center}
\resizebox{\textwidth}{!}{\includegraphics{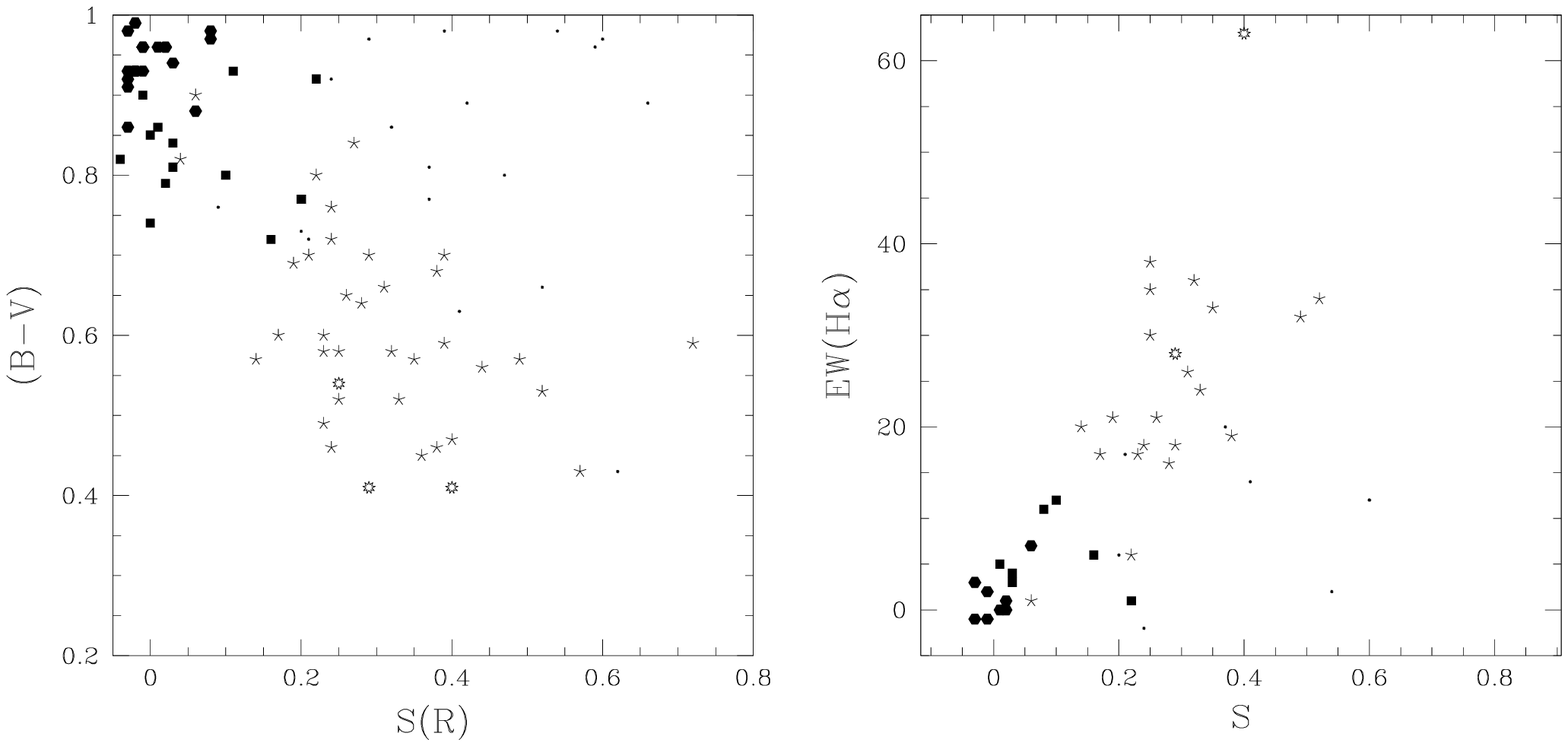}}
\end{center}
\figcaption{The correlation between the $S$ parameter and $(B-V)$ color
and H$\alpha$ equivalent width.}
\end{inlinefigure}

\begin{inlinefigure}
\begin{center}
\resizebox{\textwidth}{!}{\includegraphics{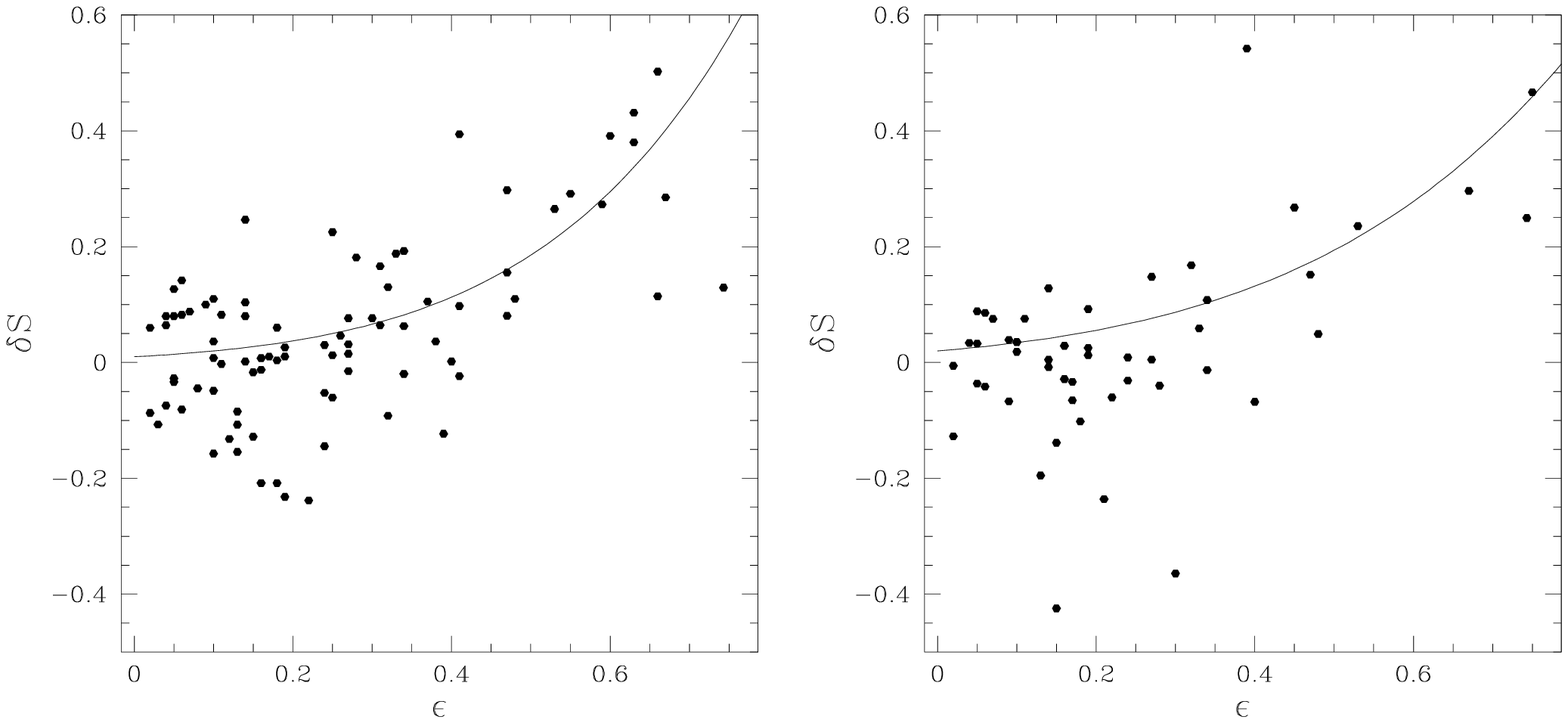}}
\end{center}
\figcaption{Residuals between the best fit between $S$ and (B-V) and
H$\alpha$ equivalent width plotted as a function of $\epsilon =$ log (a/b).}
\end{inlinefigure}

\begin{equation}
{\rm EW(H\alpha)} = (82.1\pm10) \times S' + 3.1\pm2.6,  
\end{equation}

\begin{equation}
{\rm (B-V)} = (-0.88\pm0.07) \times S' + 0.85\pm0.02.
\end{equation}

\noindent Figure~14 shows that the inclination corrected S' parameter,
measured on only a single image of a galaxy, can be 
used to determine, or at least place constraints on, the equivalent width of
the H$\alpha$ emission from galaxies and therefore its star formation
properties.  This is consistent with the clumpiness parameter $S$ as a
good measure of very recent star formation as H$\alpha$ fluxes are
a better measure of very young stars than broad-band colors (Gallagher,
Hunter \& Tutukov 1984).

\begin{inlinefigure}
\begin{center}
\resizebox{\textwidth}{!}{\includegraphics{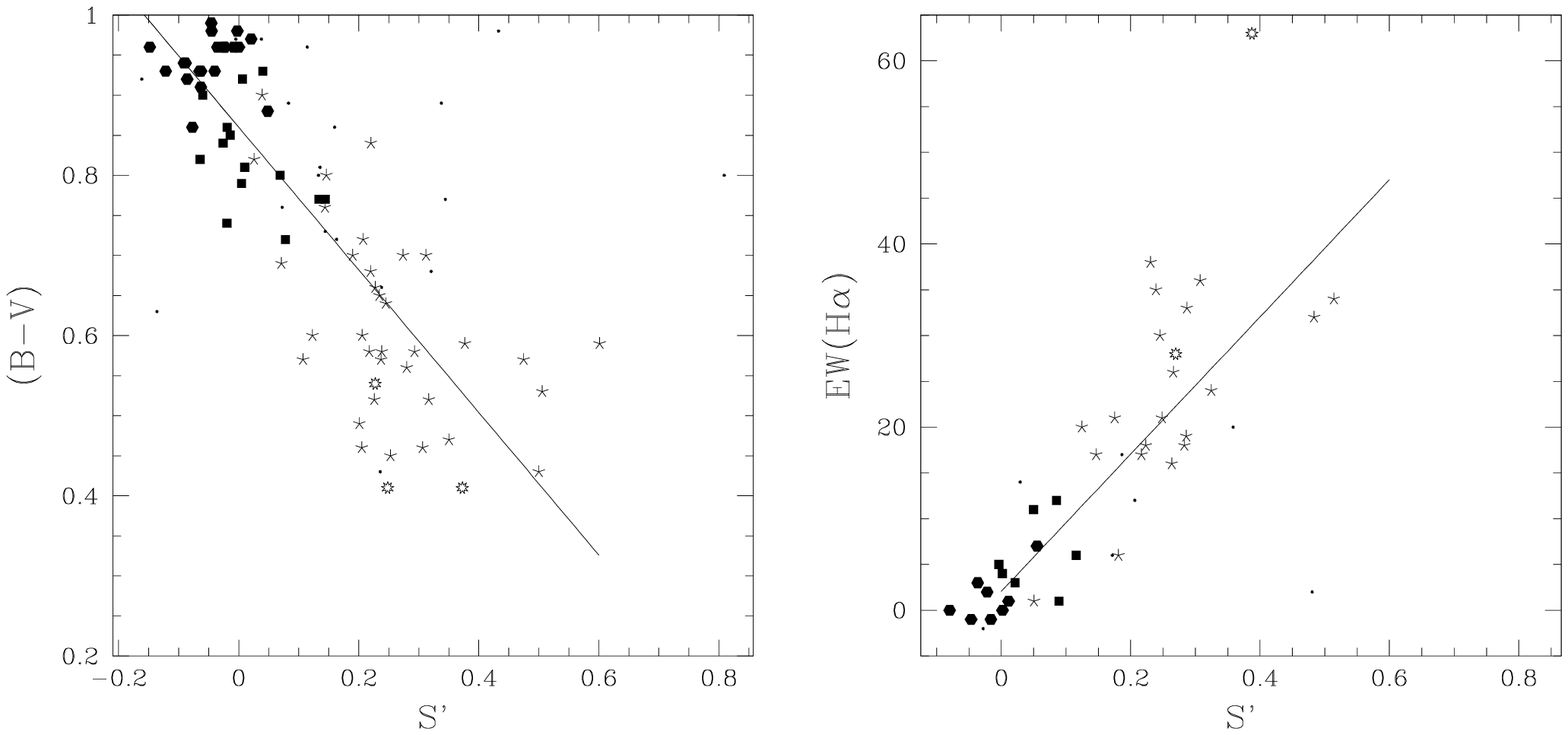}}
\end{center}
\figcaption{The relationship between the inclination corrected $S$ parameter,
S', with $(B-V)$ and H$\alpha$ equivalent width.}
\end{inlinefigure}

\section{Using the CAS system}

\subsection{Locations of Galaxies in Structural Parameter Space}

In this section we demonstrate that the CAS morphological parameters
can be used to classify all galaxy types in a three dimensional 
volume.  We also demonstrate how this system can be used
on high redshift galaxies.  A classification cube defined by the CAS values
of our sample (Tables 1 - 5) is shown in Figure~15. Each panel shows two of
the structural indexes plotted against each other, while the value
of the three parameter is denoted by the 
color of the point.  The first panel shows the concentration and asymmetry
indices where the color of the points give the value of the
clumpiness, $S$, of each galaxy.  Systems that have  $S < 0.1$
are colored red, those with $0.1<S<0.35$ green, and 
systems with $S > 0.35$ are blue.  Likewise for the $A-S$ diagram: red is for
systems with $C > 4$, green for $3<C<4$ and
blue for $C < 3$.  In the $S-C$ diagram red symbols are for
galaxies with $A<0.1$, 
green symbols are for galaxies with $0.1<A<0.35$ and blue symbols for
those with $A>0.35$.

\begin{inlinefigure}
\begin{center}
\resizebox{\textwidth}{!}{\includegraphics{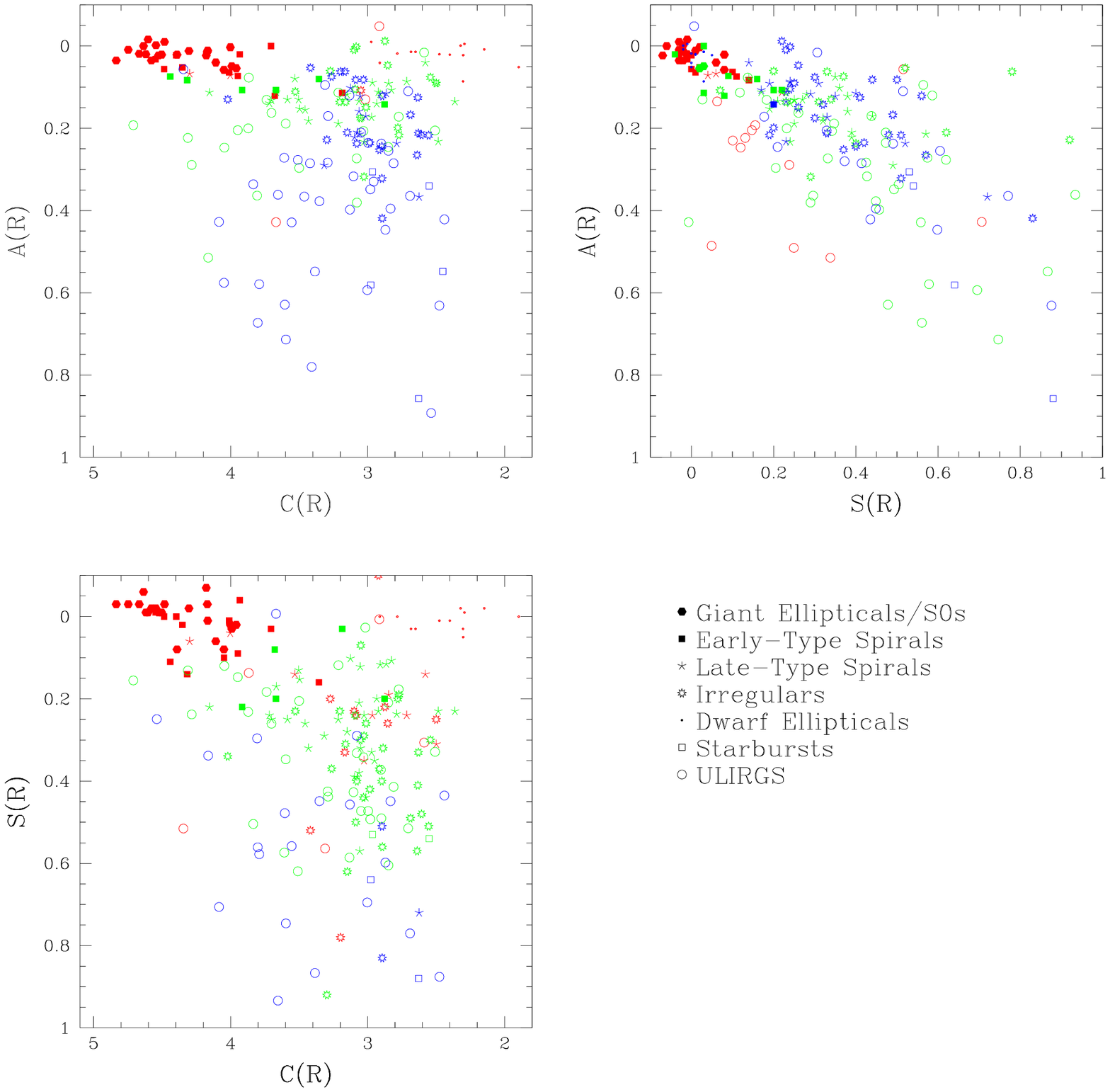}}
\end{center}
\figcaption{Three realizations of the different galaxy data sets plotted 
together by their CAS parameters.
The first panel shows the concentration and asymmetry indexes 
plotted with colored points that reflect the value of the
asymmetry for each galaxy.  Systems that have clumpiness values, $S < 0.1$
are colored red, those with $0.1<S<0.35$ are green, and 
systems with $S > 0.35$ are blue.  Likewise for the $A-S$ diagram: red for 
$C > 4$, green for $3<C<4$ and
blue for $C < 3$ and for the $S-C$ diagram: red is for
systems with $A < 0.1$, green $0.1<A<0.35$ and blue $A>0.35$.  
This figure demonstrates that when using these three morphological
parameters all known galaxy types can be distinctly separated and thus
distinguished in structural space.}
\end{inlinefigure}

The average values of the concentration,
asymmetry and clumpiness parameters, and their 1$\sigma$ variations 
are listed in Table~6 for each of
the broad galaxy types we study.  From this table and Figure~15 all major 
galaxy types are cleanly separated into various regions of CAS space.  

\subsubsection{Ellipticals}

Elliptical galaxies populate the space
of high $C$ ($<~C~> = $ 4.4$\pm$0.3), low $A$ ($<~A~> =$ 0.02$\pm$0.02), 
and low $S$ ($<~S~> = $0.00$\pm$0.04), as might be expected.  A simple 
examination of elliptical galaxy images by eye reveals that they have very 
little internal structure or asymmetries.  The light profiles of these
systems are also fairly concentrated, something that has been known since
at least Morgan (1958) and quantified by many others since then (\S 5.1).  
The physical reasons for this are quite simple -
elliptical galaxies are classified as such because they do not show evidence
for recent star formation and interactions with other galaxies.  Yet, because
they are possibly formed from a gravitational event that removes angular
momentum, or a rapid early monolithic collapse, their light is concentrated.

\subsubsection{Spirals}

Early-type spirals (Sa and Sb galaxies) have slightly higher asymmetries
($<A> =$ 0.07$\pm$0.04) and clumpiness values ($<S> = $0.08$\pm$0.08) and 
lower concentrations ($<C> = $ 3.9$\pm$0.5) than the ellipticals.
Late-type galaxies (Sc and Sd galaxies) have even higher asymmetries 
($<A> =$ 0.15$\pm$0.06) and clumpiness values 
($<S> = $0.29$\pm$0.13) and lower light concentrations
($<C> = $ 3.1$\pm$0.4).  This is also what we might have expected since spiral 
galaxies
are known to contain modest star formation and slight asymmetries
produced by star formation or minor interactions with other galaxies (CBJ00).  
These galaxies have lower light concentration
values than the giant ellipticals because of their appreciable disks
containing angular momentum.

\subsubsection{Dwarf Irregulars}

Dwarf irregulars have similar asymmetries as the late-type disks 
($<A>$ = 0.17$\pm$0.10), but have higher clumpiness values 
($<S>$ = 0.40$\pm$0.20) and lower concentrations ($<C>$ = 2.9$\pm$0.3).
This is further consistent with the idea that lower-mass Hubble type
galaxies are dominated by star formation.  The fact that the average
concentration index is low for these dwarf irregulars is also consistent
with this index telling us about galaxy formation processes, and the
scale of a galaxy, as these are all faint systems with M$_{\rm B} < -18$ 
(van Zee 2000).

\subsubsection{Dwarf Ellipticals}

The above results are not too surprising for galaxies on the Hubble
sequence whose concentration and asymmetry parameters have previously
been described (e.g., CBJ00).  The interesting properties
of Figure~15 are for the objects
considered unusual or outside the Hubble sequence that give the real power to 
the CAS structural indices for distinguishing galaxies in different phases
of evolution. 

The dwarf ellipticals in our sample have
low asymmetries ($<A>$ = 0.02$\pm$0.03) 
and low clumpiness values ($<S>$ = 0.00$\pm$0.06), similar to
the giant ellipticals (Table~6), which is probably why they were thought 
to be of a similar class.  Yet dwarf ellipticals differ significantly from 
the giant 
ellipticals in terms of their light concentrations (Figure~15; Table~6) 
revealing a difference in scale, and possibly formation history, from the giant
ellipticals (\S 5.1).   The dwarf ellipticals
have an average concentration value $<C> = 2.5\pm0.3$, which has no
overlap with the giant ellipticals, with $<C> = 4.4\pm0.3$, different 
at a formal significance of $\sim$ 6$\sigma$.  This demonstrates the 
fundamental difference of scale reflected in the light concentrations of these
galaxies. 

\subsubsection{Mergers}

The interacting/merging galaxies in our sample, which includes some  
starbursts (SB) (Markarian
8, NGC 3690, and NGC 7673, see Table 2) and the ULIRGs (Table 5), have
on average the highest asymmetry and clumpiness values ($<A>_{\rm SB}$ = 
0.53$\pm$0.22, $<S>_{\rm SB}$ = 0.74$\pm$0.25 and 
$<A>_{\rm ULIRGs} = 0.32\pm0.19$, $<S>_{\rm ULIRGs}$ = 0.50$\pm$0.40).
These are also the only galaxy types in our sample with
high asymmetry and clumpiness parameters, with values 
larger than $S > 0.4$ and $A > 0.35$, except a few irregulars with
$A > 0.35$ and a number with $S > 0.4$.  This is clearly due to the active
evolution occurring in these galaxies.  Both starbursts and ULIRGs have
intense star formation which results in high clumpiness 
values.  Many of these galaxy types are in an active stage of
a major merger, and thus have high asymmetry values, as argued
earlier in Conselice et al. (2000b).  

The concentration values for
these two populations, is interestingly, not different from Hubble
sequence galaxies with average concentrations, $<C>_{\rm SB}$ = 2.7$\pm$0.2
and $<C>_{\rm ULIRGs}$ = 3.5$\pm$0.7.  This is consistent with violent 
relaxation occurring in ULIRGs, in which concentrated light profiles are 
established rapidly 
(e.g., Lynden-Bell 1967; Wright et al. 1990).

From these arguments, and the average values listed in Table~6, and individual
galaxies plotted in Figure~15, we argue that galaxies in
generally agreed upon different phases of galaxy evolution can be
differentiated in a structural-morphological space defined by the 
concentration, asymmetry, and clumpiness indices.   In the next section
we show that the CAS parameters can be measured reliably at various
S/N ratios, resolutions, and redshifts therefore making the CAS system a 
powerful one for understanding the evolution of galaxy populations 
from $z \sim 3$ until today.   

\subsection{Usefulness of the CAS system at High Redshift}

As we are interested in applying the CAS structural parameters in future 
papers to decipher 
evolution from low to high$-z$, it is
natural to ask how these parameters behave as galaxies
become less resolved and fainter due to cosmological effects.  To address
this question we simulate nearby bright galaxies, used in this paper, as
to how they would appear at various higher redshifts as observed in 
deep Hubble Space Telescope images, namely the WFPC2 and NICMOS 
Hubble Deep Field North and the ACS GOODS fields.  After simulating
these galaxies we remeasure their CAS values and compare the derived
parameters to their $z \sim 0$ values.
  These simulations are done
using the approach of Giavalisco et al. (1996), although we use 
a more general prescription described below. 

\subsubsection{Method}

When simulating an image of a galaxy at $z_1$ to how it would appear 
at $z_2$, with $z_2 > z_1$, several factors must be considered.  The first is 
the re-binning factor, b, which is the reduction in apparent size of
a galaxy's image when viewed at higher redshift.
The other factors are the relative amounts of flux from the sky and galaxy
and the noise produced from the galaxy, sky, dark current and imaging
instrument (e.g., read noise).
To calculate these, we generalize the procedure for simulating a hypothetical 
galaxy with actual size $d$ (measured in
kpc, for example) imaged at a redshift 
$z_{1}$ which we want to simulate at $z_{2}$.  
At a given higher redshift $z_2$ this galaxy will subtend an angle,

\begin{equation}
\theta_{z_2} = \frac{d}{A_{z_2}} = d \frac{(1+z_2)^{2}}{L_{z_2}},
\end{equation}

\noindent where $A_{z}$ is the angular diameter distance, which is related
to the luminosity distance ($L_{z}$) by $A_{z} = L_{z}/(1+z)^{2}$.  The 
number of pixels in a galaxy, as imaged at a redshift $z$, is given by 
$n_{z} = \theta_{z}/s_{z}$ where $s_{z}$ is the angular size of the pixels
used in the detector to image the galaxy at redshift $z$.  At the redshift 
$z_{2}$ the galaxy
will sub-tend an angle $\theta_{z_2}$ which will appear on the
simulated image with a size in pixels $n_{z_1}/n_{z_2}$ times smaller than 
the original image with an angular size $\theta_{z_1}$.  The 
ratio of these galaxy sizes in pixel units is the binning factor (b) which 
can be written as

\begin{equation}
b = n_{z_1}/n_{z_2} = \frac{\theta_{z_1}}{\theta_{z_2}} \frac{s_{z_2}}{s_{z_1}} = \frac{(1+z_{1})^{2}}{(1+z_{2})^{2}} \frac{L_{z_2}}{L_{z_1}} \frac{s_{z_2}}{s_{z_1}}.
\end{equation}

\noindent The luminosity distance, $L_{z}$ can be expressed as an analytic
formula for cosmologies with $\Omega_{\lambda}$ = 0 (Giavalisco
et al. 1996), but for the cosmology used here, $\Omega_{\lambda}$ = 0.7, 
$\Omega_{\rm m}$ = 0.3,
$L_{z}$ must be computed numerically (e.g., Peebles 1980).
The amount the observed galaxy must be reduced in surface brightness
is computed by using the conservation of energy, that is
the luminosity before and after reshifting is the same, or

\begin{equation}
4\pi \alpha_{z_1} N_{z_1} p_{z_1} L_{z_1}^{2} (1+z_{1}) = 4\pi \alpha_{z_2} N_{z_2} p_{z_2} L_{z_2}^{2} (1+z_{2}) \frac{\Delta \lambda_{z_1}}{\Delta \lambda_{z_2}},
\end{equation}

\noindent for a galaxy observed in filters at central rest frame
wavelengths $\lambda_{z_1}$ and $\lambda_{z_2}$ and filter widths of 
$\Delta \lambda_{z_1}$ and $\Delta \lambda_{z_1}$. $N_{z}$ is the total number
of pixels covering a galaxy at $z$ and p$_{z}$ is the average ADU s$^{-1}$ 
per pixel.   The calibration constant
$\alpha_{z}$ is in units of erg s$^{-1}$ cm$^{-2}$ \AA$^{-1}$ 
(ADU s$^{-1}$)$^{-1}$.
Since $N_{z} \sim (\theta_{z}/s_{z})^{2}$ and by using
eq. (12) we can rewrite eq. (13) in terms of $p_{z_2}$ in ADU counts as

\begin{equation}
p_{z_2} = p_{z_1} \frac{\alpha_{z_1}}{\alpha_{z_2}} \left(\frac{s_{z_2}}{s_{z_1}}\right)^{2} \left(\frac{1+z_{1}}{1+z_{2}}\right)^{4} \frac{\Delta \lambda_{z_2}}{\Delta \lambda_{z_1}} \frac{t_{z_2}}{t_{z_1}}
\end{equation}

\noindent where $t_{z}$ is the respective exposure times at the observed
redshift $z_{1}$ and simulated higher redshift 
$z_{2}$.  An extra factor of (1+$z_{2}$)/(1+$z_{1}$) has
been applied to eq. (14) to account for the smaller rest-frame
wavelength observed at $z_{1}$ than at $z_{2}$.  The observed flux of the
simulated galaxy as seen at $z_2$ in ADU units can be computed through
the use of eq. (14).  Note that ideally we minimize k-corrections in this 
approach by matching the observed wavelengths at $z_{1}$ and $z_{2}$.

The sky background is then added to these images by the addition of 
$B_{z} \times t_{z}$, where $B_{z}$ is the amount of light from the background
in units of ADU s$^{-1}$.  Note that the original background in the $z_{1}$
image has already been removed prior to artificially redshifting.  Noise
components are then added in, which include: read-noise scaled for the number
of read-outs, dark current and photon noise from the background.  Finally
the resulting image is smoothed by a PSF appropriate for various HST
instruments as generated by Tiny-Tim (Kriss et al. 2001).    

To understand
how well we can measure the CAS parameters at high redshift we simulated
various galaxies in our sample out to redshifts of $z = 3$ using the above
approach.  As we use rest-frame optical images we implicitly
assume a zero morphological k-correction.  Although galaxies clearly appear
different in the rest frame optical and ultraviolet (e.g., Windhorst 
et al. 2002; cf. Conselice et al. 2002c for starbursts) we want to isolate 
the effects of
decreased resolution, and increased noise, on the CAS parameters so therefore
use the same image.

To understand and quantify how well the CAS parameters can be measured
at high redshifts we carried out three different simulations.  Two of
the simulations consisted of placing the Frei sample at
redshifts from $z \sim 0.5$ to 3 as they would appear in the Hubble Deep
Field WFPC2 F814W (I) and the GOODS F850L (z) images.  We then measure 
the CAS values on these simulated images to determine how they have 
changed. 

\begin{inlinefigure}
\begin{center}
\resizebox{\textwidth}{!}{\includegraphics{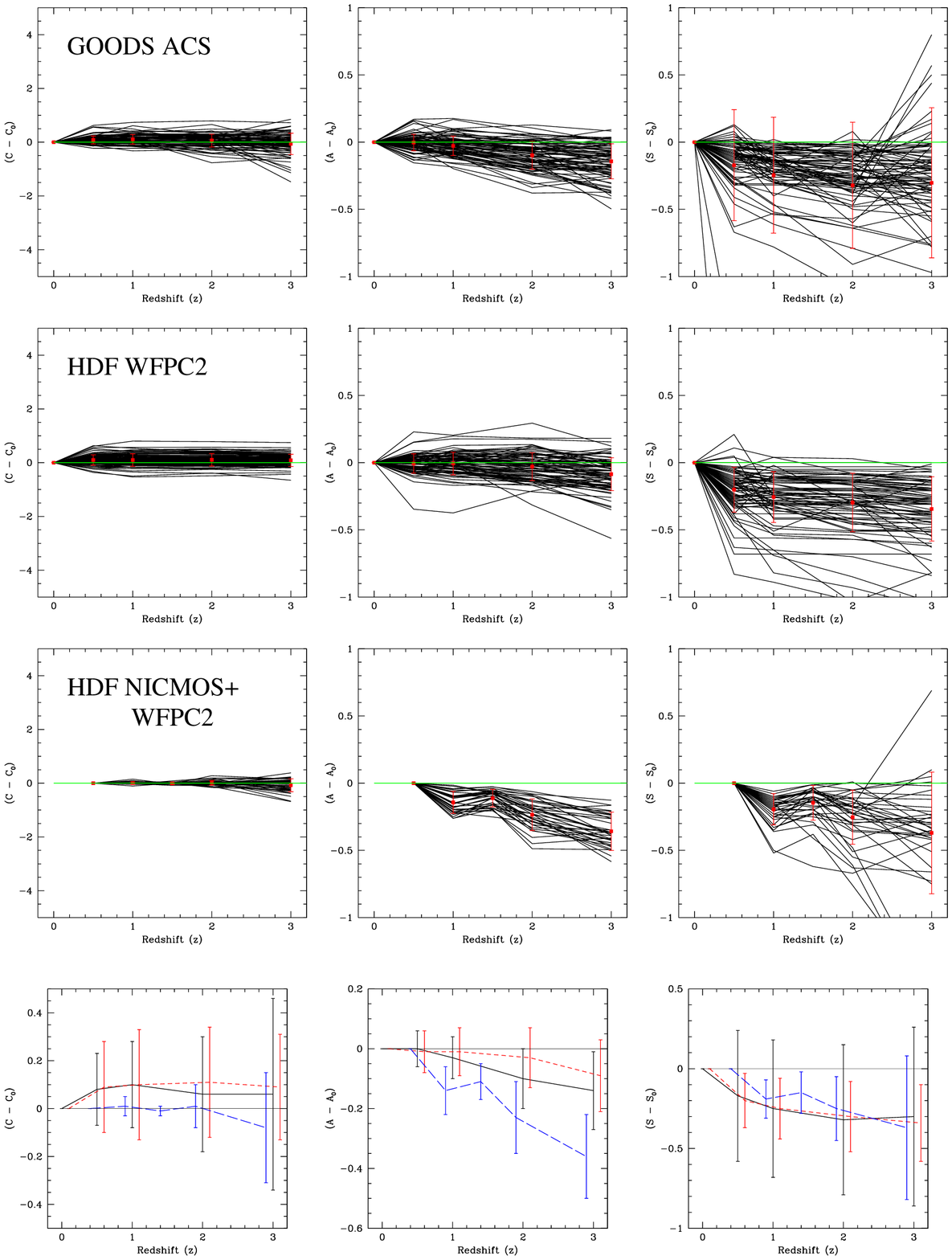}}
\end{center}
\figcaption{Determination of redshift effects on the measurement of the CAS
parameters.  In the top two panels (a and b) the 82 galaxies that comprise
the Lowell Frei sample (see Frei et al. 1996; Conselice et al. 2000a) are
simulated to how they would appear at z = 0.5,1,2 and 3 in (a) the GOODS
ACS F850L (z) imaging and (b) the F814W (I) image of the Hubble Deep Field 
North.  The third panel (c) shows how the 38 galaxies with M$_{\rm B} < -18$
in the Hubble Deep Field North would appear at higher redshifts in the
rest-frame B-band.  This is done by using the WFPC2 + NICMOS images  of
the Hubble Deep Field North, which allows the rest-frame appearance of
galaxies to be seen out to $z \sim 2.5$. We simulate this $z \sim 0.5$
sample to $z = 1, 1.5, 2$ and 3.
The numbers plotted are with reference to the
$z \sim 0$ (or $z \sim 0.5$ value for NICMOS) CAS parameters (C$_{0}$, A$_{0}$,
S$_{0}$). The green line shows the zero level and the red dot and 
errorbars are the averages and 1$\sigma$ variations at each simulated 
redshift.  The bottom panel shows the average differences and their
1 $\sigma$ scatter values for the GOODS (solid dark line), HDF WFPC2 
(red dashed line) and the NICMOS+WFPC2 simulations (long blue dashed line).  }
\end{inlinefigure}

We also performed a
simulation accounting for morphological
k-corrections by simulating galaxies with M$_{\rm B} < -18$ found in
the Hubble Deep Field North between $0.4 < z < 0.7$ to higher 
redshifts.  At $z \sim 1.2$ the rest-frame B-band morphologies of these
galaxies are sampled in the observed NIR, thus we carry out simulations
using the WFPC2 images for these galaxies at $z < 1.2$
and the NICMOS HDF-N observational parameters in the F110W and F160W bands 
(Dickinson et al. 1998) at $z > 1.2$.

\subsubsection{GOODS ACS Fields}

The GOODS fields consist of deep BViz imaging in the Hubble Deep Field
North and Chandra Deep Field South regions taken as part of the initial ACS 
Treasury program (see http://www.stsci.edu/ftp/science/goods/).  For our 
simulations we use
the observing parameters for the F850L (z-band) observations 
(Giavalisco et al. 2003 in prep).  Our simulated images therefore match the 
conditions in which these galaxies are observed, sans
morphological k-corrections as we use the rest-frame B-band image of
each galaxy in the simulation.  After simulating the 82 Lowell Frei
galaxies as to how they would appear in the z-band image of ACS GOODS, we
run the CAS program on each system.  This is first done
by remeasuring the 1.5 $\times (\eta = 0.2$) radii, and then measuring the 
CAS parameters within
this.  We do not assume a radius for each galaxy, but
remeasure everything as if it were a new galaxy, with 
unknown $z = 0$ properties.

The three panels in Figure~16a show how the concentration 
index $C$, asymmetry index $A$, and clumpiness index, $S$, change for
each of the 82 galaxies as a 
function of redshift with respect to their $z = 0$ values, with
$z = 0.5, 1, 2$, and 3 simulated as described above. 
The average differences from $z = 0$ and their 1$\sigma$ deviations are 
plotted as red dots and error-bars, and are listed in Table~7.    In
Figure~16d the average differences for the GOODS CAS parameters and their 
1$\sigma$ variations are plotted as the solid dark line.

\subsubsection{WFPC2 Hubble Deep Field North}

We perform a similar simulation as done for the GOODS z-band field, but using
the observing parameters in which the WFPC2 F814W Hubble Deep Field
observations were acquired (see Williams et al. 1996).  The results of
these simulations are shown in Figure~16b, and are plotted in Figure~16d
as the dashed red line.  The same procedure described in \S 6.2.2 is also used
to simulate these galaxies.

\subsubsection{NICMOS+WFPC2 HDF - Eliminating the Morphological k-correction}

By simulating galaxies into the NICMOS images of the Hubble Deep Field North 
taken by
Dickinson et al. (1998), combined with matched seeing
WFPC2 images from Williams et al. (1996) we can
determine the rest-frame morphological appearance of galaxies from $z = 0$ to
$\sim 2.5$.  This eliminates the morphological k-correction, as the NICMOS 
F110W (1.1 $\mu$m) 
and F160W (1.6 $\mu$m) bands allow us to sample rest-frame B-band
morphologies out to $z \sim 2.5$.
From this we can determine how morphological indices
are changing purely as a function of redshift, without worrying about 
band shifting effects.
A more detailed analysis of this problem is presented in Conselice
et al. (2003b).   We carried out this simulation by
using low redshift galaxies in the HDF, and simulating them
to higher redshifts.  For this, we used the 38 galaxies with 
M$_{\rm B} < -18$ at redshifts $0.4 < z < 0.7$ found in the HDF North.
The result of these simulations are shown in Figure~16c when simulating
the $z \sim 0.5$ galaxies to $z = 1,1.5,2$ and $3$. The
average CAS value differences between the values at $z \sim 0.5$ and
the higher redshifts are shown in Figure~16d as a blue line.

\subsubsection{Simulation Results}

From these diagrams it is clear that at least the asymmetry
and concentration indices are reproducible out to $z = 3$, within a
reasonable scatter.  This is especially the case when using the average 
differences for a given population, as it is nearly zero, or can
be corrected to zero, for galaxies
at any redshift, the possible exception being the NICMOS observations which
have a lower resolution than the WFPC2 or ACS data.  

The asymmetry and clumpiness spatial indices also 
decline, on average, as a function of redshift while the concentration index
slightly increases from a zero difference as a function of redshift.
From Figure~16 it is
clear that one cannot be certain after measuring CAS parameters on a single
high redshift galaxy what its $z \sim 0$ value would be.  The average
CAS values of a given population can however be corrected to a statically
average
$z \sim 0$ value which can be used for comparing evolution.  The CAS values
of a single galaxy at $z > 1$ will, on average, have an uncertainly that
will make their use somewhat limited until deeper high resolution imaging
is carried out.

These simulations are also generally idealized and do not include the improved
image quality and effective resolution produced by drizzling.
The average differences and their
1$\sigma$ scatter also improve when only considering very bright
galaxies, or those with active star formation.  A detailed analysis of
this, and how CAS values for different galaxy populations change with
redshift in these simulations is however
beyond the scope this paper, but will be address fully in future studies
of high redshift systems.

\section{Summary}

One of the major problems in extragalactic astronomy is the absence of
a well defined physical classification system for galaxies.  Ideal
classification systems are those that classify objects according to their
most salient physical or evolutionary properties and processes.  In this 
paper we
introduce a model independent morphological system that measures fundamental
formation
processes and their past history and uses these features to classify 
all galaxies.  This is done
by using the concentration index ($C$), the asymmetry parameter ($A$) and a
clumpiness index ($S$).  

The concentration index reveals how concentrated a galaxy's
light is, while the asymmetry index and clumpiness indices quantify how 
disturbed
and clumpy light distributions are.  We argue in this paper that the
concentration index is a measure of the relative fraction of light in
bulge and disk components, i.e., the B/T ratio, and is a measure of the scale
of a galaxy - its size, luminosity and mass.  Concentration thus reveals the 
past formation history of a galaxy.  The asymmetry index is shown
to be good at identifying galaxies which are currently undergoing major 
mergers. The clumpiness index is found to correlate very well with the 
H$\alpha$ equivalent width of galaxies, and to a lesser degree with the 
integrated $(B-V)$ color of a galaxy.  

After constructing a volume out of the CAS parameters for 240 nearby
galaxies in all phases of evolution, we find that all major nearby galaxy 
types (ellipticals, late and early-type 
spirals, irregulars, starbursts, mergers, and dwarf ellipticals) can
be cleanly distinguished from each other automatically and computationally.
We also demonstrate that this classification system can be applied at
high redshift by showing that average CAS parameters can be accurately
determined for any galaxy population seen in  
deep Hubble Space Telescope images out to $z \sim 3$.   As the 
Advanced Camera for Surveys will provide an unprecedented morphological
view of high redshift galaxies, we hope that the CAS system will be useful
for understanding and determining how galaxy evolution and formation has
occurred.

I thank my collaborators, especially Richard Ellis, Matt Bershady, Mark 
Dickinson, Jay Gallagher and Rosie Wyse whose work with me led 
to the present ideas and paper.  Richard Ellis, Andrew Benson, Matt Bershady,
Tommaso Treu, and an anonymous referee made valuable suggestions after 
reading various versions of 
this paper. I also thank Eric Peng for allowing me to use his GALFIT program.
This work was supported by a National Science Foundation Astronomy
and Astrophysics Fellowship and by NASA
HST Archival Researcher grant HST-AR-09533.04-A, both to CJC.

\clearpage
\appendix

\section{Deconstructing the Hubble Sequence}

The Hubble morphological system has been 
the dominate galaxy classification paradigm since its first publication
in 1926\footnote{Hubble (1926) was the first to set forth
the systematic classification system that now bears his name, although
early photographic work on galaxies reveled that 
spiral galaxies come into many forms, including the
barred or $\phi$ classification (Curtis 1918).  Several other
galaxy classification systems predate Hubble's, such as Wolf's
(1908) rich descriptive taxonomy, although these classification
were not generally in use after 1945 (Sandage 1975).} (Hubble 1926).   
The basic tuning-fork morphological
system was later expanded and revised (Hubble 1936; Sandage 1961; 
de Vaucouleurs 1959; van den Bergh 1960) to include additional
galaxies such as the Sd and S0 types as well as morphological
features such as rings, `S' shapes,  and other
features.  These later morphological properties however add 
little to the physical meaning of a `morphological-type' and are
rarely used in practice expect for very specific purposes.

Despite the wide acceptance of the Hubble system, there are many well know
problems with it, for example see van den Bergh (1998), 
Conselice et al. (2000a), van den Bergh, Cohen \& Crabbe (2001), and
Abraham \& van den Bergh (2001) for
discussions of this problem.   One major issue is that at
redshifts higher than $z \sim 1$ the Hubble sequence no longer provides a 
useful frame-work for understanding galaxies (van den Bergh et al. 2001). Not
only is it difficult, to impossible, to accurately obtain a Hubble type from
an algorithm (Naim et al. 1995a; Odewahn et al. 2002) but classification
experts often do not agree among themselves on subjective 
classifications (Naim et al. 1995b).  The result of this is that the
physical meaning of a Hubble
type is unclear (e.g., Roberts \& Haynes 1988) as a Hubble type does not
alone uniquely reveal any physical property.   That is, the Hubble 
classification 
system does not identify, or distinguish between, galaxies involved 
in very different modes of evolution,
except for the basic trichotomy between ellipticals, spirals, and
irregulars which vary significantly between each other in most
measurable properties. 

Hubble classifications are also internally inconsistent within its
own classification requirements.  
This can be seen by investigating the criteria for classification into
different Hubble types.  According to Sandage (1975), the criteria for 
classification on the Hubble sequence relies on three subjective
properties:~``(1) the 
size of the nuclear bulge relative to the flattened disk, (2) the character of
the spiral arms, and (3) the degree of resolution into stars and HII regions
of the arms and/or disk.'' 
This is similar to the original criteria used by Hubble (1926) 50 years
earlier: ``(1) relative size of the the unresolved nuclear region; (2)  
extent to which the arms are unwound; (3) degree of resolution in the
arms.''  Criterion (3) is entirely insufficient for
comparing galaxies, since the degree of resolution depends
strongly on distance, telescope resolving power and seeing, and we argue has 
rarely been a real criteria for
placing galaxies on the Hubble sequence (cf. Sandage \& Tammann 1981).  
Criterion (2) relates to the pitch 
angle of spiral arms from the nuclear or bulge region, and to the presence 
of strong bars in spirals, while criterion (1) is 
the relationship between spiral bulge and disk sizes.  

Since criteria (3) is a largely inadequate for classifying galaxies at
different resolutions
let us examine the other two conditions to determine how well they correlate.
To answer this question, Kennicutt (1981) measured the pitch angles ($\psi$) of
113 spiral galaxies to determine how $\psi$ values correlate with
Hubble types as classified in the Revised Shapley-Ames Catalog (Sandage
\& Tammann 1981).
As shown by Kennicutt (1981) there is only a weak correlation between
the Hubble type and pitch angle of spiral arms.    We can reexamine
Kennicutt's (1981) data to determine how well pitch angles
correlate with T-types as determined by de Vaucouleurs et al. (1993) where
each spiral T-type is based on the Sandage (1961) criteria.  Figure~17 shows 
this comparison.  Although 
on average pitch angles increase for
latter T-types, the 1$\sigma$ variations of the average $\psi$ for
each T-type clearly overlap for galaxies later than,
and including, Sbs.    This is an indication that the
methodology for classifying Hubble types is not sound, as the
criteria is neither internally consistent, nor,
as discussed above, does it reveal unique galaxy populations.
As such, classifiers using the Hubble sequence
often must pick either estimating the bulge to disk ratio, or the pitch angle 
of spiral arms, to place a galaxy into a type, as has been
recognized for many years (e.g., Sandage 1975). 

\begin{figure}
\vskip 1in
\plotfiddle{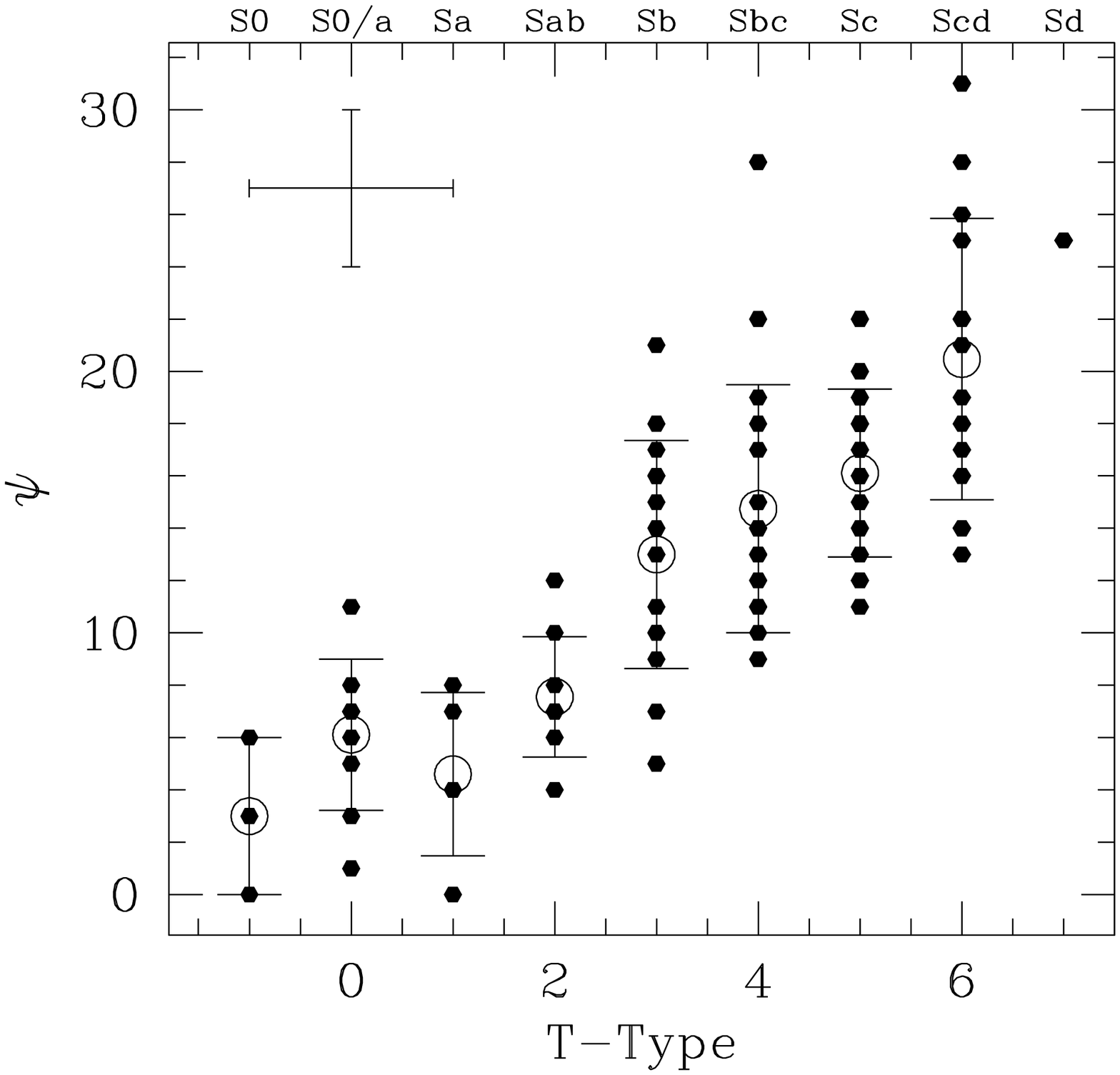}{6.0in}{0}{80}{80}{-250}{-70}
\vskip -1in
\caption{The pitch angle $\psi$ for spiral galaxies, as measured by 
Kennicutt (1981), plotted against
Hubble type estimates as determined by de Vaucouleurs et al. (1993).
The open circle denotes
the average pitch angle at each T-type and the plotted error is the
1$\sigma$ variation.  There is no statistically
significant difference between the measured pitch angle for galaxies
classified as Sb (T=3) or later.}
\end{figure}

Furthermore, recent results show that the ratio of bulge and disk sizes is 
scale free along the Hubble Sequence (e.g., de Jong 1996; Coureau, de Jong \& 
Broeils 1996); that is,
the ratio of bulge to disk scale lengths is roughly constant for 
different spiral Hubble types. It is also often 
assumed that the Hubble sequence correlates with physical
features, such as stellar populations, colors and star formation histories, 
since the work of Holmberg (1958). As for pitch angles, this is true for 
Hubble types in the mean, 
but there is a large scatter of physical properties at every Hubble stage
(e.g., Roberts \& Haynes 1994; Kennicutt 1998).  There is also  
evidence that all spiral Hubble classes have roughly the same average star 
formation
properties (e.g., Devereux \& Hameed 1997; Hameed \& 
Devereux 1999).  Therefore, a basic Hubble type does not reveal any 
fundamental physical information about a particular galaxy, and
in practice a Hubble type is often used only as a guide for understanding
a galaxy rather than as a useful physical description.   This is the
result of the most prominent
features of galaxies being not necessarily useful for determining the
physical evolution of a galaxy, as we argue below.

\subsection{Bars, Rings and Spiral Arms}

Galaxy bars are very visible, and as such,
have been a major classification component since the work of Curtis
(1918).  Hubble (1926) made the distinction between barred and unbarred
spirals the major division in spiral galaxies, and bars put the `fork'
in the Hubble tuning fork.  De Vaucouleurs (1959) considered bars important 
enough that his classification system uses weak or strong bars as a 
fundamental criterion.

Although the classification of spiral galaxies into barred and unbarred
is often done, it is important to ask if this is a useful feature
for understanding galaxies and their evolution.  Many disk galaxies, if not
most, have some kind of central bar (Buta \& Combes 1996).  This is 
especially true when examining galaxies at red wavelengths
(e.g., Eskridge et al. 2000) where
the bar fraction can be over 60\% in the near infrared. Galaxies with or
without bars do not differ significantly in terms of physical features
such as stellar populations (de Vaucouleurs 1961) which suggests that bars 
cannot be a fundamental feature, in themselves.  Bars are not seen in other 
galaxy types such as ellipticals and therefore are only properties of disks.  
In principle, bars 
can also form and reform without changing the basic evolution of a galaxy.
They are still important however as they trace the effects, and are the
causes of evolutionary phenomenon, such as interactions
and star formation.

The rings and s-shaped patterns popular in the de Vaucouleurs (1959) 
revision of the Hubble Sequence and used by Sandage (1961) and others, are 
related to bars in some way and are also not salient classification features
for similar reasons.   Rings are useful for
identifying the locations of resonances in galaxies (Buta \& Combes 1996), and
again are likely transient in nature.  There are also no obvious physical
differences between galaxies that have bars and rings and those that do
not (van den Bergh 1998).  Rings can, like bars, also 
harbor star formation (Buta
\& Combes 1996) and are likely related to bars and spiral arms 
(e.g., Buta \& Combes 1996).

Like bars and rings, spiral patterns
are a transient phenomenon  and trace the location of galactic disks.    
They are also
probably the most spectacular morphological feature in galaxies and
are the principle feature of the Hubble sequence.
As traditionally observed at optical wavelengths, most spiral arms contain
bright young stars that only account for a very small fraction of a galaxy's
mass.  Spiral arms also look very different when observed at various
wavelengths, as first discussed by Zwicky (1957).  At shorter wavelengths,
particularly in the near infrared, pitch angles decrease and nuclear
regions increase in relative size, changing the Hubble type.  The near 
infrared is in fact a better place to study stars in galaxies since
a better idea of the true mass distribution is
obtained (e.g., Block \& Puerari 1999).  The fact that bulge to disk
ratios, pitch angles, and spiral arm structure changes with  
wavelength is also a hint that the properties of spiral
arms are not ideal criteria for classifying galaxies.
In summary, bars, rings and spiral arms are all features that
signify internal dynamics of disk galaxies, but they alone are not
fundamental evolutionary changes in a galaxy, and therefore 
should not be used to physically classify galaxies.

\section{Theoretical Classification}

Galaxy evolution can be understood through the use of stellar
structures in a variety of models, such as in the hierarchical context or 
monolithic collapse, by the simple use of scaling relationships
between quantities and observables, and an understanding of the various
components and evolutionary processes occurring in galaxies.  To demonstrate 
this we
start by writing the mass of a galaxy as a function of time, M$_{\rm T}(t)$, 
as a composition of stars, gas and dark matter as,

\begin{equation}
{M_{\rm T}(t)} = {M_{\rm S}(t)} + {M_{\rm G}(t)} + {M_{\rm DM}(t)},
\end{equation}

\noindent where M$_{\rm S}(t)$, M$_{\rm G}(t)$ and M$_{\rm DM}(t)$ are the 
stellar, gaseous and dark matter mass of a galaxy at a given time, $t$. 
In the CDM scenario, at $t_{0}$=0 when the dark matter
halo first collapses, M$_{\rm S}(t_{0})$ = 0.   Over time, star formation 
occurs and various forms of mass will be accreted and removed from a galaxy 
due to interactions and mergers, changing the composition, and
quantity, of a galaxy's mass.  At a time $T > t_{0}$, the total amount
of mass added to, or removed from, a galaxy by interactions, mergers
and accretion, $\delta M_{\rm T}(T)$ is given by,

\begin{equation}
\delta M_{\rm T}{(T)} = \delta {\rm M_{S}}(T) + \delta {\rm M_{G}}(T) + \delta {\rm M_{DM}}(T) = \int^{T}_{0} \dot{M}_{\rm T}(t) {\rm d}t = \int^{T}_{0} \dot{M}_{\rm S}(t) {\rm d}t + \int^{T}_{0} \dot{M}_{\rm G}(t) {\rm d}t + \int^{T}_{0} \dot{M}_{\rm DM}(t) {\rm d}t,
\end{equation}

\noindent where $\dot{M}_{\rm T}(t)$ is the mass accretion or removal rate in
units of \solm time$^{-1}$, and $T$ is the age of the galaxy when observed.  
We further divide the mass flux into stellar $\dot{M}_{\rm S}(t)$, gaseous 
$\dot{M}_{\rm G}(t)$ and dark matter $\dot{M}_{\rm DM}(t)$ components. 

In this scenario gas will be
obtained by two methods, in addition to any gas present at initial formation
(M$_{\rm G,0}$).  Gas can be obtained through gravitational processes such
as mergers\footnote{In this
instance mergers implies that the gas is brought into a galaxy in discrete
units, such as attached to an existing galaxy or dark matter halo.
This is opposite to gas accreted from the intergalactic medium which
is not necessarily added to the galaxy in discrete units, but is added smoothly
over time.}, $\delta M_{\rm G,G}(t)$, 
and from gas accreted smoothly over time through deposition, 
$\delta M_{\rm G,D}(t)$.  We can then write the addition of a galaxy's gas 
mass as 
two components, such that $\delta M_{\rm G} = \delta M_{\rm G,G} + 
\delta M_{\rm G,D}$.   The total gas mass can be written as, 
$M_{\rm G}(t) = M_{\rm G,G}(t) + M_{\rm G,D}(t) + M_{\rm G, 0}(t)$.
Gas in a galaxy will be converted into stars as a function of
time, $t$, by a conversion rate, $\alpha(t)$, defined as the fraction of
gas mass converted into stars per unit time, such that the star formation,
$\delta M_{\rm G \rightarrow S}$, is given by

\begin{equation}
\delta M_{\rm G \rightarrow S}(T) = \int^{T}_{0} \alpha(t)M_{\rm G}(t) {\rm d}t = \int^{T}_{0} \alpha(t)(M_{\rm G,G}(t) + M_{\rm G,D}(t) + M_{\rm G,0}(t)) {\rm d}t = \delta M_{\rm S,G}(T) + \delta M_{\rm S,D}(T) + \delta M_{\rm S,0}(T),
\end{equation}

\noindent where $M_{\rm G}(t)$ is the total gas mass as a function of time, 
and $M_{\rm G,G}$, $M_{\rm G,D}$ and $M_{\rm G,0}$ are 
the gas masses of a galaxy obtained from through the gravitational process of 
mergers,  gas deposited from accretion, and the original gas present.

The stellar mass of a galaxy, $M_{\rm S}(t)$, is therefore formed from gas
placed into a galaxy by several physical methods. It could have 
formed through smooth gas deposition (M$_{\rm S,D}$),  gravitational collapse 
(i.e., mergers) of existing stars (M$_{\rm S,G})$, or conversion from
the original gas in the galaxy, $M_{\rm S,0}$,
such that $M_{\rm S}(t)= M_{\rm S,G}(t) +  M_{\rm S,D}(t) + M_{\rm S,0}(t)$. 
More generally, we can write the total 
stellar, gas, and dark matter masses of a galaxy as a function of time as

\begin{equation}
M_{\rm S}(T) = \int^{T}_{0} \left(\dot{M}_{\rm S}(t) + \alpha(t)M_{\rm G}(t)\right) {\rm dt},
\end{equation}

\begin{equation}
M_{\rm G}(T) = {\rm M_{G,0}} + \int^{T}_{0} \left(\dot{M}_{\rm G}(t) - \alpha(t)M_{\rm G}(t)\right) {\rm dt},
\end{equation}

\begin{equation}
M_{\rm DM}(T) = {\rm M_{DM,0}} + \int^{T}_{0} \left(\dot{M}_{\rm DM}(t)\right) {\rm dt}.
\end{equation}

\noindent Where the total mass of a galaxy at a given
time is given by equation B1.
Equation B6 assumes that the dark matter in a galaxy is non-baryonic,
or at least cannot be converted into stars.
We also do not account here for stellar feedback into the intergalactic
medium, which would include a term in eq. B5 to account for supernova material
and other mass ejecta from stars, but this can be easily done (e.g., Sandage
1986).

Most of the original stellar
mass in galaxies probably formed from an initial deposition of gas, thus
$M_{\rm S}(t_{\rm gal}) =  
M_{\rm S,D}(t_{\rm gal})$ where $t_{\rm gal}$ is 
the time when the first generation of stars formed.  Later, through mergers 
and star formation processes, the total stellar mass of a galaxy will increase
and stellar mass will be redistributed through gravitational processes, if
mergers occur, while
new stars are formed from gas deposition and gas obtained through mergers.
We can relate the time variable ratio of mass created through these various
methods by the parameter $D(T)$, defined as the fraction of stellar mass
formed through gravitational processes:

\begin{equation}
D(T) = \frac{M_{\rm S,G}(T)}{M_{\rm S}(T)} = \frac{\int^{T}_{0} \left(\dot{M}_{\rm S}(t) + \alpha(t) M_{\rm G,G}(t)\right){\rm dt}}{\int^{T}_{0} \alpha(t) M_{\rm G}(t) {\rm dt}}.
\end{equation}

\noindent This fraction can then be used, knowing the 
value of $D(T)$,
to track the gross morphological evolution of a galaxy's stellar and
how it is globally distributed.     

In principle the CAS parameters can be represented
by three observable terms in equations B1 - B7.  Light concentration, which 
is the fraction of stars produced through gas depositionally placed, can be
represented by $D(T)$, which does not change with time unless star formation,
accretion, or mergers occur.
The asymmetry parameter can be represented by $M_{\rm S,G}(t)$ 
and the clumpiness, $S$, which is
a measure of star formation, can likewise
be represented by $\delta M_{\rm G \rightarrow S}(t)$ (eq. B3). 

Understanding the correspondence between these theoretical quantities and
measurable properties requires a very good knowledge of the internal
evolution of galaxies, including stellar evolution and 
dynamical relaxation processes and their related time scales.  We can however
create other dimensionless parameters, similar to B7, that in principal 
directly correlate with the asymmetry and clumpiness, indicating the 
evolutionary state
of a galaxy.  As this paper is focused on answering the question of whether
or not we can use stellar light to measure these properties, the 
integrated deposition $D(T)$ (eq. B7), star formation $F(T)$ and 
interaction, $I(T)$ features weighted by mass and time can be written
in terms of measurable dimensionless indices as:

\begin{equation}
F(T) = \int^{T}_{0} \frac{t}{T} \frac{\alpha(t) M_{\rm G}(t)}{M_{\rm S}(T)} {\rm dt}.
\end{equation}

\begin{equation}
I(T) = \int^{T}_{0} \frac{t}{T} \frac{\left|\dot{M}_{\rm S}(t)\right|}{M_{\rm S}(T)} {\rm dt}
\end{equation}

\noindent These three parameters (eq. B7 - B9) are dimensionless  physical 
morphological quantities that correlate with the stellar appearance of a 
galaxy and fundamental evolutionary processes, such that $C~\alpha~D$, 
$A~\alpha~I$, and $S~\alpha~F$.  Equations (B8) and (B9) 
are weighted by time since older star formation and interactions will
have less of an effect on the appearance of a galaxy.  Likewise,
galaxies with higher stellar masses will have appearances that in general 
are less effected by star formation, and stars lost or gained from 
galaxy interactions.  The parameter $D(T)$ can of course change, 
but it is not an integral of time since the amount of stellar mass settled 
by depositional or gravitational processes does not passively evolve without 
outside influences.  Star formation in a galaxy, or
interactions between galaxies however does morphologically evolve 
in a few Gyrs (i.e., stellar aging, cluster dissolution and 
relaxation processes smooth out effects from star formation and
galaxy interactions).

\subsection{Predicting Structural Features}

By using equations (B7) through (B9), it is possible to investigate the 
theoretical evolution a galaxy's structure.  This requires understanding some
basic physical features such as the fraction of gas mass converted into stars
per unit time, $alpha(t)$, and the history of interactions with other galaxies.

There are many ways to investigate simple cases of theoretical morphology
and its evolution, only a few of which we examine here.
A simple case is to assume that a galaxy contains an initial 
amount of gas that declines exponentially at a constant scale-factor, 
$\alpha$, due to star formation. This gas
mass is converted into stellar mass, thereby producing star 
formation that will induce morphological changes into the system under
study. Solving equation (B5)
using a constant $\alpha$, with no new additions of gas, gives an exponential 
decrease in the gas mass,

\begin{equation}
M_{\rm G}({\rm t}) = {\rm M_{G,0}}\times {\rm exp}(-\alpha~t).
\end{equation}

\noindent By using eq. (B10) with eq. (B8), the index $F(t)$ can be written
in this situation as,

\begin{equation}
F(T) = \frac{{\rm M_{G,0}}}{T \times M_{\rm S}(T) \alpha}\times(1 - {\rm exp}(-\alpha T) \times (\alpha T + 1)).
\end{equation}

\noindent The evolution of the $F(T)$ parameter as a function of redshift
in this scenario is
plotted in Figure 18a and b for a $\Omega = 1$ universe.  Figure 18a shows 
seven different curves for various
values of $\alpha$ from 0.3 for the highest curve, to $\alpha = 0.9$ for the 
curve with the lowest values at the higher redshifts.  Figure 18b shows 
$F(t)$, as a function of redshift, when
$\alpha$ is held constant at 0.2, while allowing
the ratio M$_{\rm 0,G}$/M$_{\rm S}$ to vary from 1 to 0.4.  
We can likewise write the interaction parameter, $I(t)$, in a similar way, 
assuming that the amount of mass gained or
lost declines exponentially with time such that $|\dot{M}_{\rm S}(t)| = 
|\dot{M}_{0} \times {\rm exp}(-\beta t)|$, as

\begin{equation}
I(T) = \frac{|{\rm \dot{M}_{0}}|}{T \times M_{\rm S}(T) \beta^{2}}\times(1 - {\rm exp}(-\beta T) \times (\beta T + 1)).
\end{equation}

\noindent Figure~19 plots I($T$) in this form,  
where the parameter $\beta$ is varied between 0.1 - 0.7.   Figure~19b
shows I($T$) when varying the ratio of $|{\rm \dot{M}_{0}}|/{\rm M_{S}}(T)$.
Both Figure~18 and 19
demonstrate that if galaxies have exponentially declining star formation and
interaction histories then
their morphologies are dominated by these effects at redshifts from $z = 1$ 
to 3.  This is broadly consistent with observations that show galaxies
in this redshift range have blue colors and high star formation rates
(e.g., Ellis 1997).

Another simple star formation and galaxy interaction 
scenario that can be investigated is when the amount of gas available for 
star formation, $M_{\rm G}$, and
mass removal/addition rate, $|\dot{M}_{\rm S}(t)|$ and $\alpha$, are
constant as a function of time.  This would be the case in the
situation where instantaneous recycling of gas occurs after star formation, or
gas is continuously accreted onto a galaxy.
In these cases, the parameters $F(T)$ and $I(T)$ reduce to,

\begin{equation}
F(T) = \frac{T}{2} \frac{\alpha \times M_{\rm G}}{M_{\rm S}(T)},
\end{equation}

\begin{equation}
I(T) = \frac{T}{2} \frac{|\dot{M}_{\rm S}|}{M_{\rm S}(T)},
\end{equation}

\noindent the forms of which are plotted in Figure 18c and 19c for the same 
seven ratios of M$_{\rm G,0}$/M$_{\rm S}$ and 
$|{\rm \dot{M}_{S}}|/{\rm M_{S}(T)}$ used in Figure~18b and 19b.

The third and final form we investigate are star formation and 
interaction events occurring in random `bursts' during galaxy evolution.  
Figure 18d and 19d show the result of this scenario, 
where the spikes are at the redshifts when star formation and mergers
occur for 10 Myrs, and then turn off.   We can solve for both $F(t)$ and 
$I(t)$ when  both star formation and interactions occur 
at these discrete times.  The form of these equations are,

\begin{equation}
F(T) = \frac{\alpha}{T \times M_{\rm s}(T)} \sum_{i=0}^{N} t_{i} \times M_{G}(t)_{i} \Delta T_{i},
\end{equation}

\begin{equation}
I(T) = \frac{1}{T \times M_{\rm s}(T)} \sum_{i=0}^{N} t_{i} \times |\dot{M}_{S}(t)|_{i} \Delta T_{i},
\end{equation}

\noindent where $N$ is the number of star burst or merger events, $t_{i}$ is 
the time when each event occurs, and $\Delta T_{i}$ is the time duration of 
each event.
These simple scenarios are meant to show that if we knew the detailed 
interaction and star formation history of each
galaxy, we could use eqs. (B7 - B9) to calculate its gross morphological
appearance, and the theoretical values of $C$, $A$ and $S$, as a function of 
redshift.

\begin{figure}
\plotfiddle{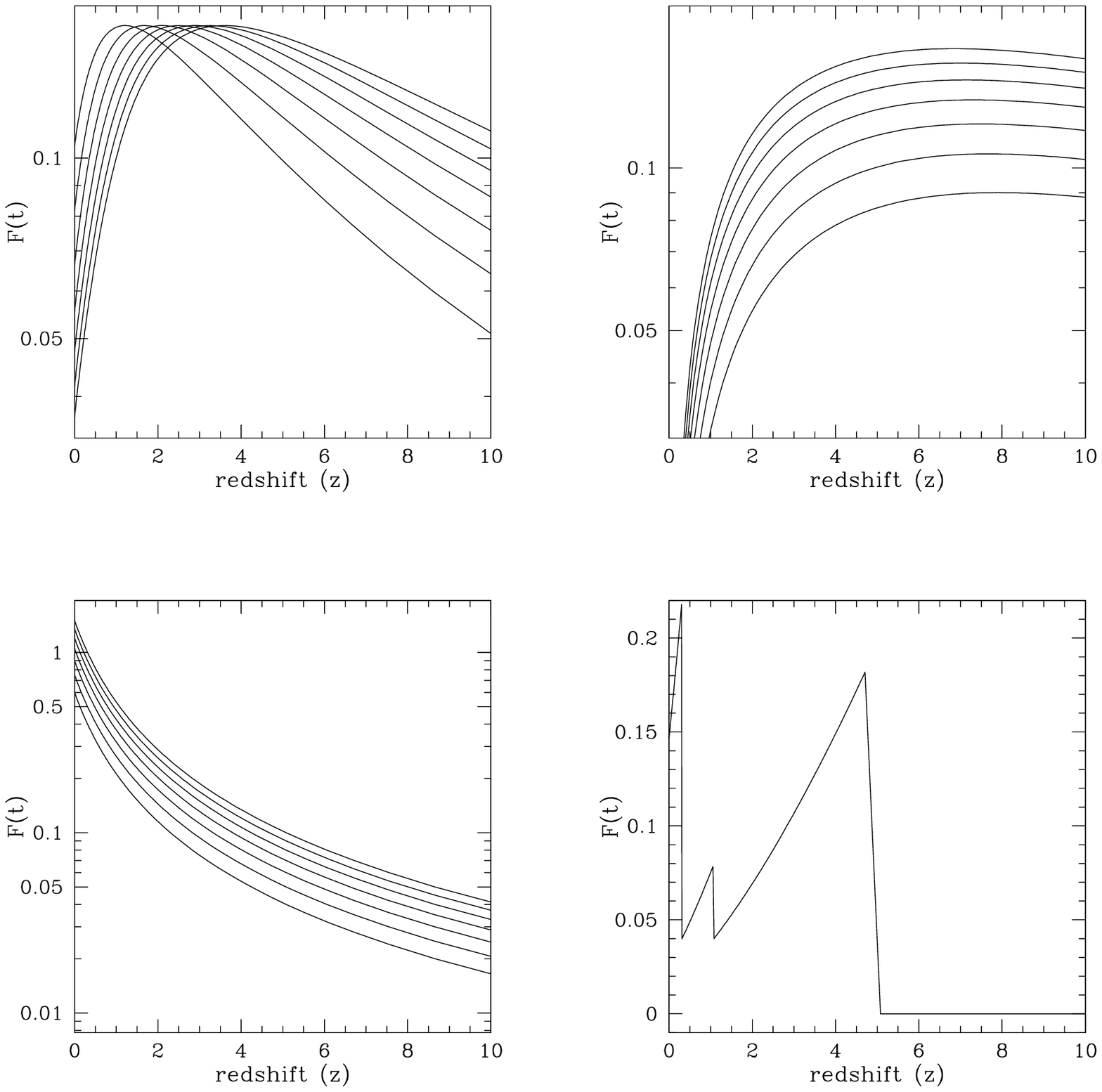}{6.0in}{0}{80}{80}{-250}{-70}
\vskip 0in
\caption{How the $F$ parameter is predicted to change as a function of
redshift in different galaxy evolution scenarios (see text).}
\end{figure}

\begin{figure}
\plotfiddle{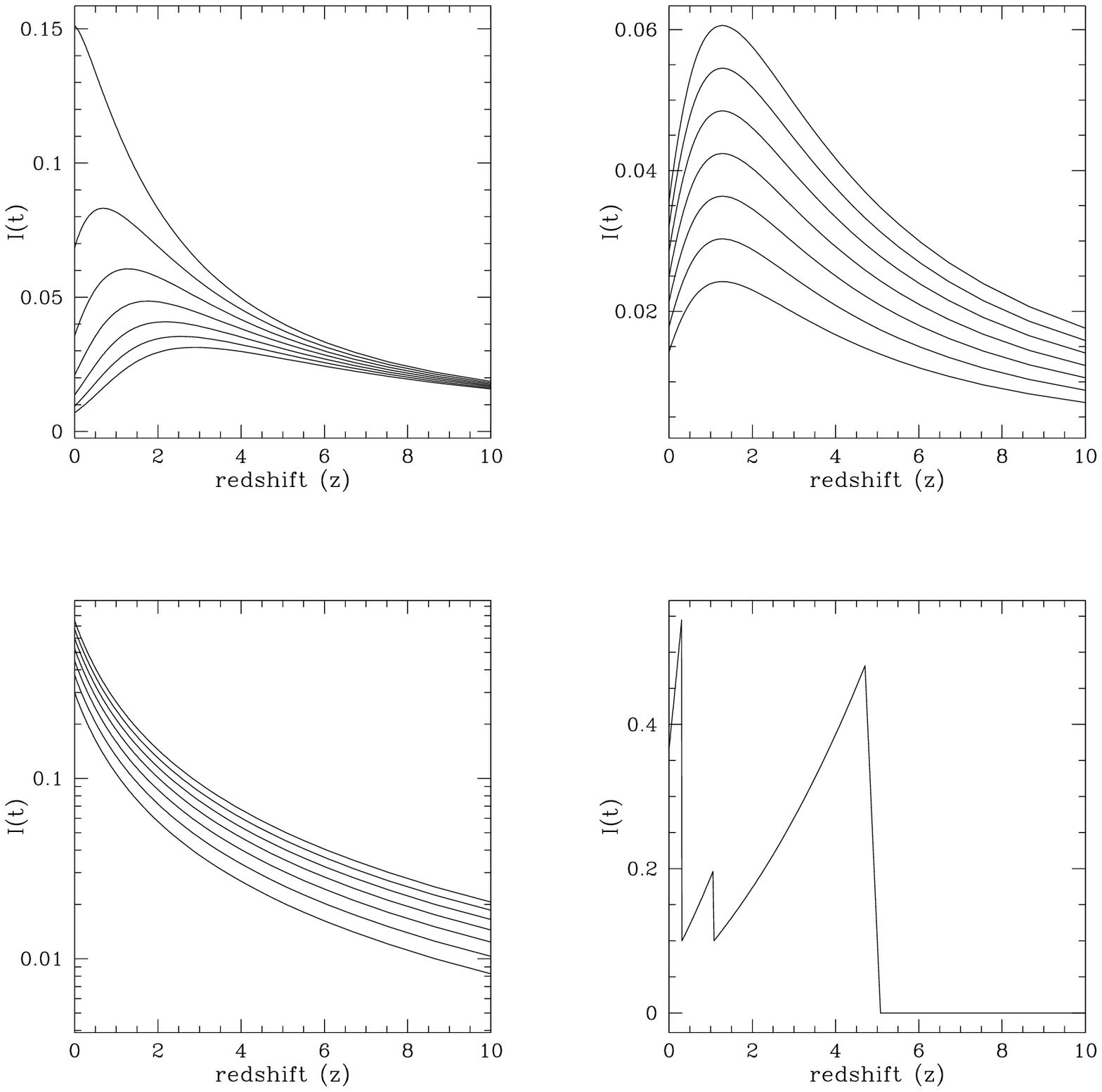}{6.0in}{0}{80}{80}{-250}{-70}
\vskip 0in
\caption{How the $I$ parameter is predicted to change as a function of
redshift in different galaxy evolution scenarios (see text).}
\end{figure}

\clearpage

\begin{figure}
\plotfiddle{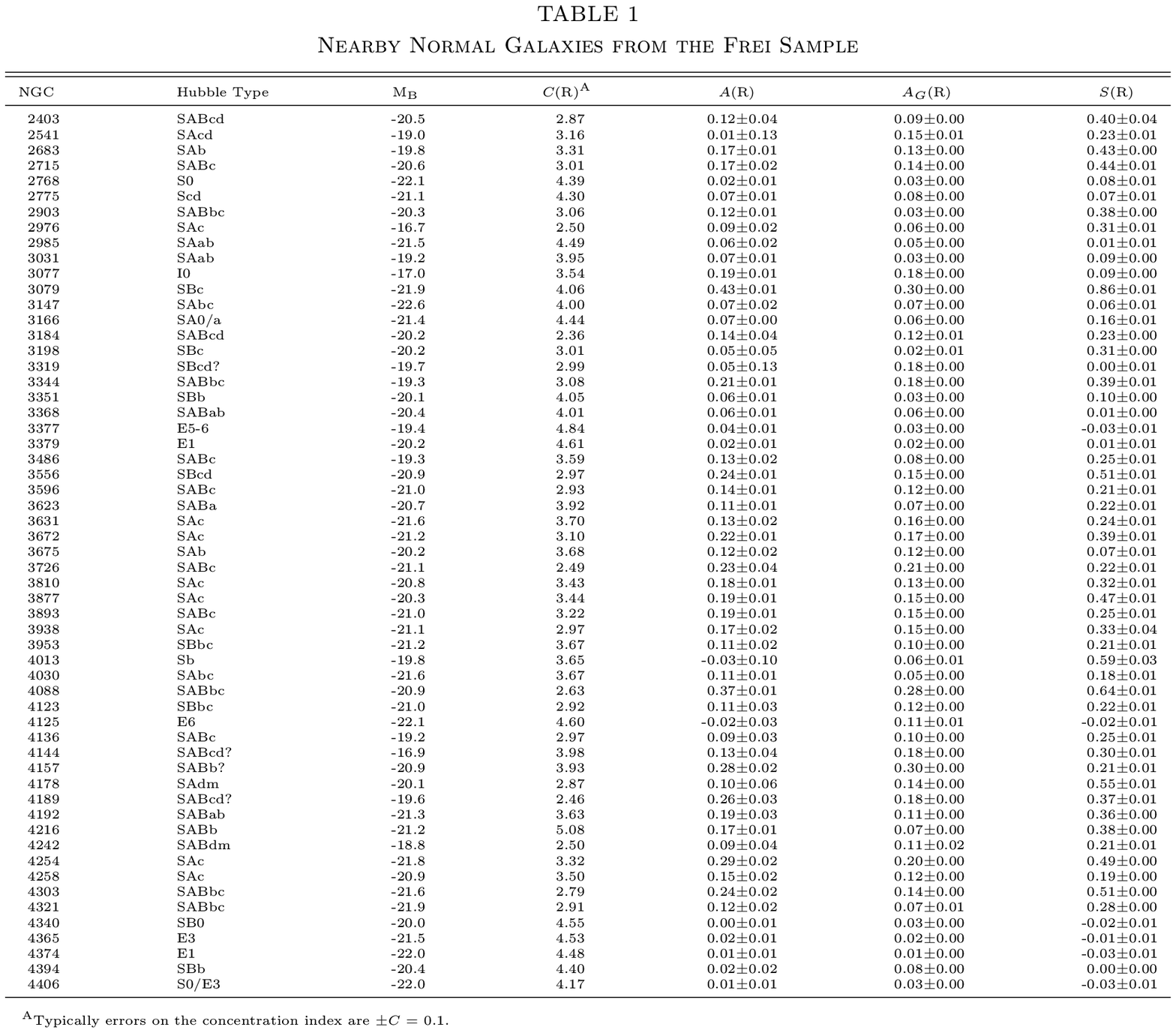}{6.0in}{0}{100}{100}{-310}{-170}
\end{figure}

\begin{figure}
\plotfiddle{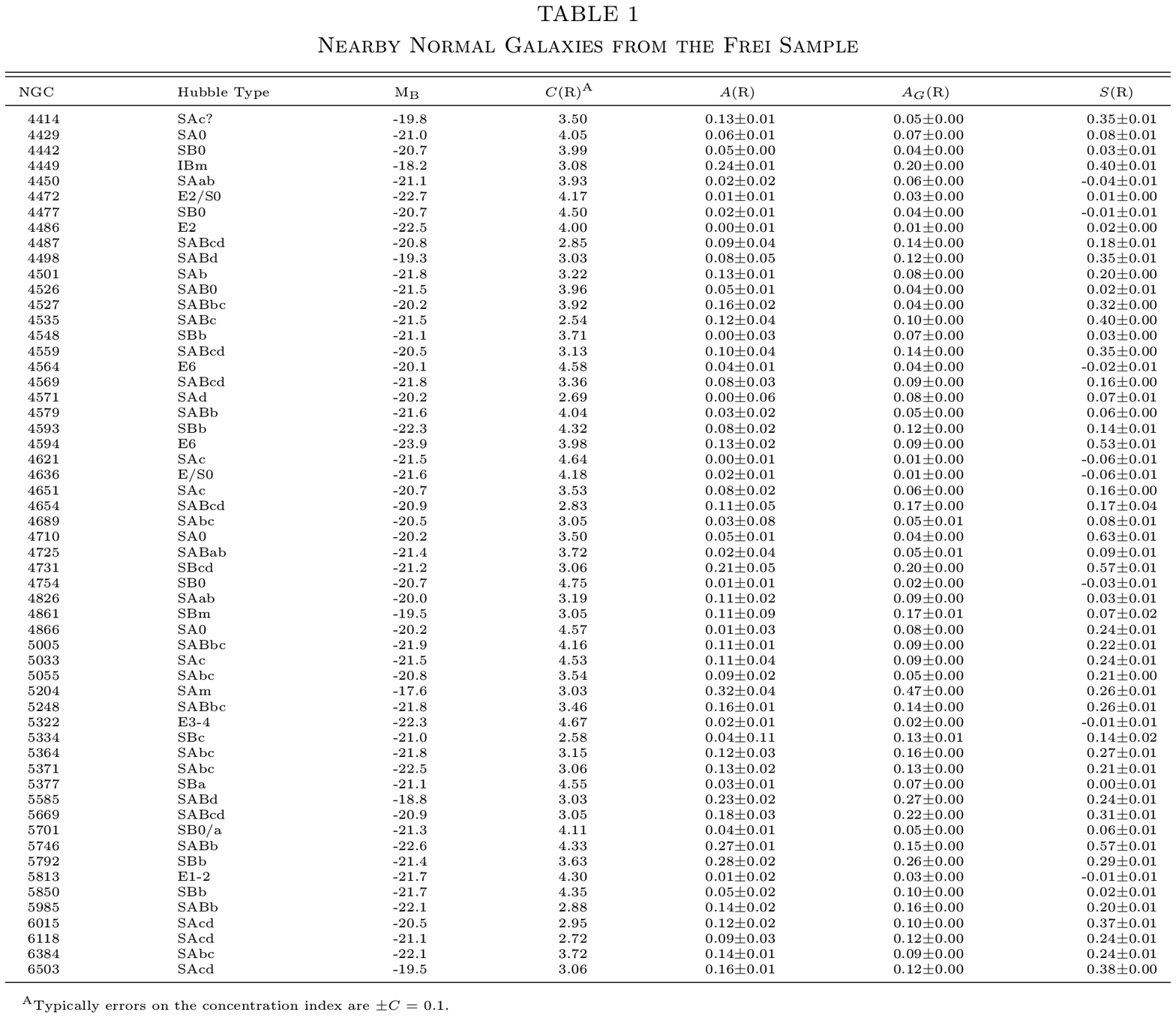}{6.0in}{0}{100}{100}{-310}{-170}
\end{figure}
\newpage

\begin{figure}
\plotfiddle{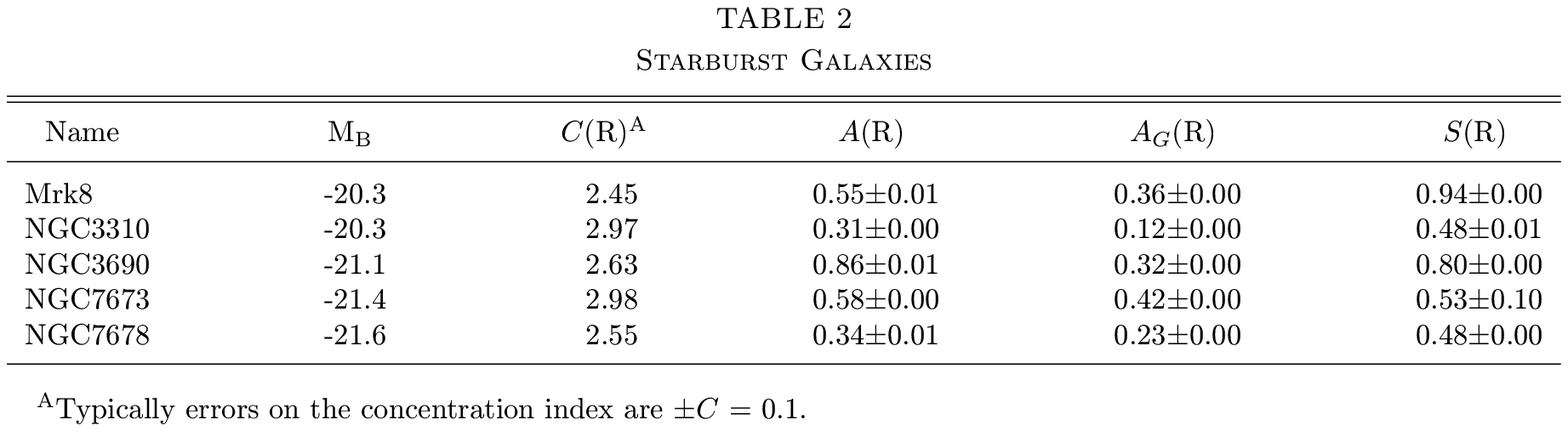}{6.0in}{0}{100}{100}{-310}{-170}
\end{figure}

\begin{figure}
\plotfiddle{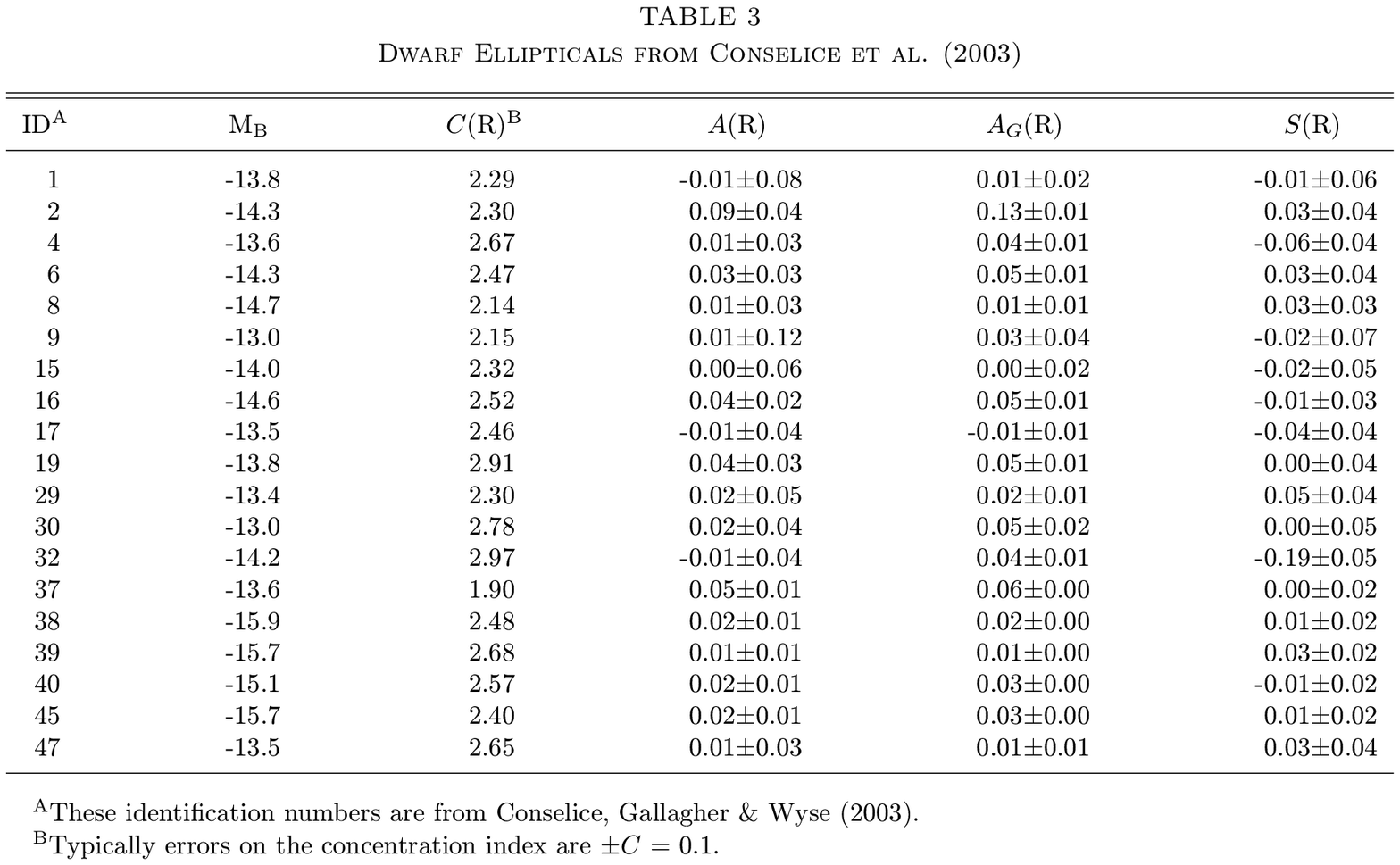}{6.0in}{0}{100}{100}{-310}{-170}
\end{figure}

\begin{figure}
\plotfiddle{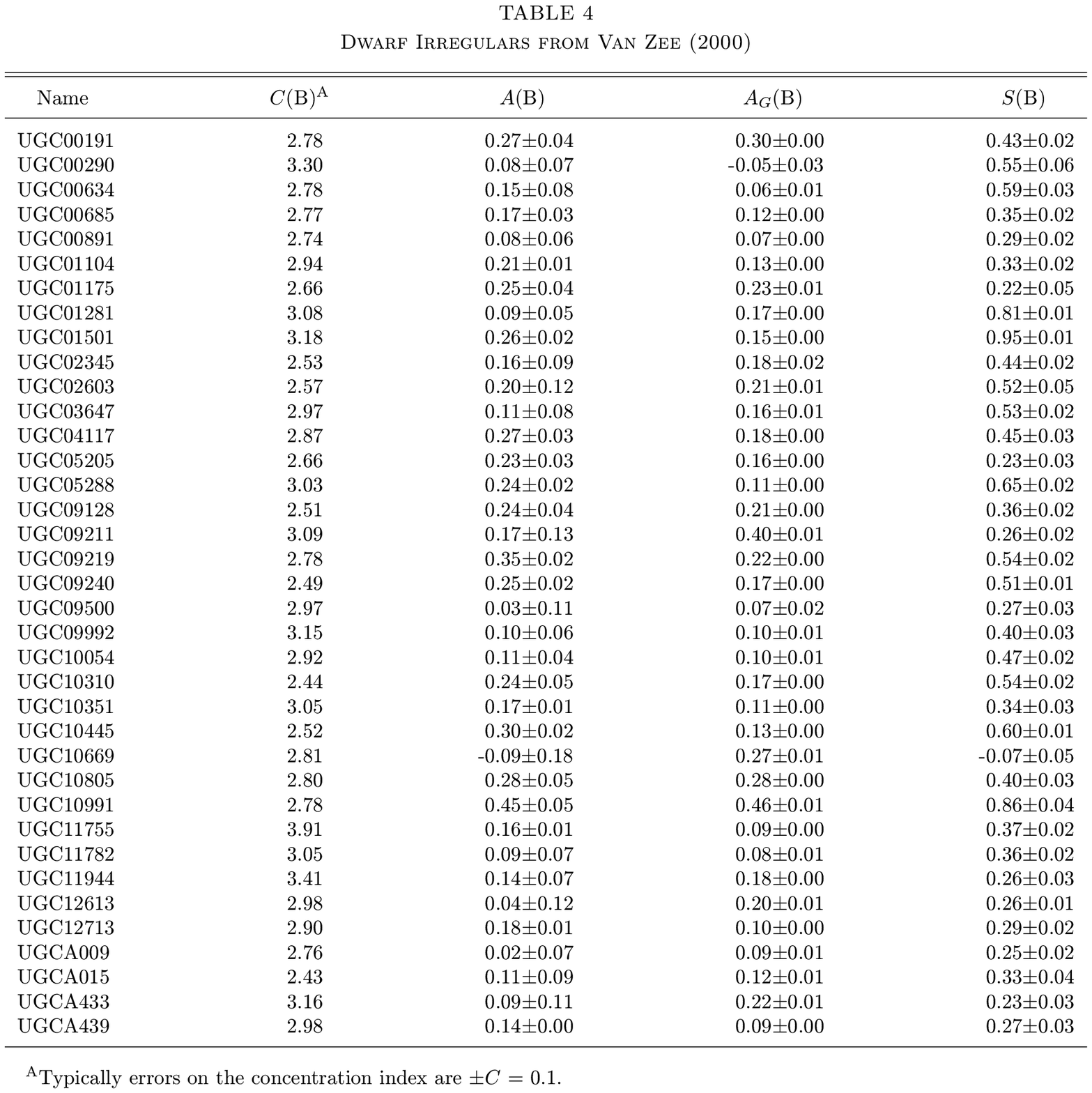}{6.0in}{0}{100}{100}{-310}{-170}
\end{figure}

\begin{figure}
\plotfiddle{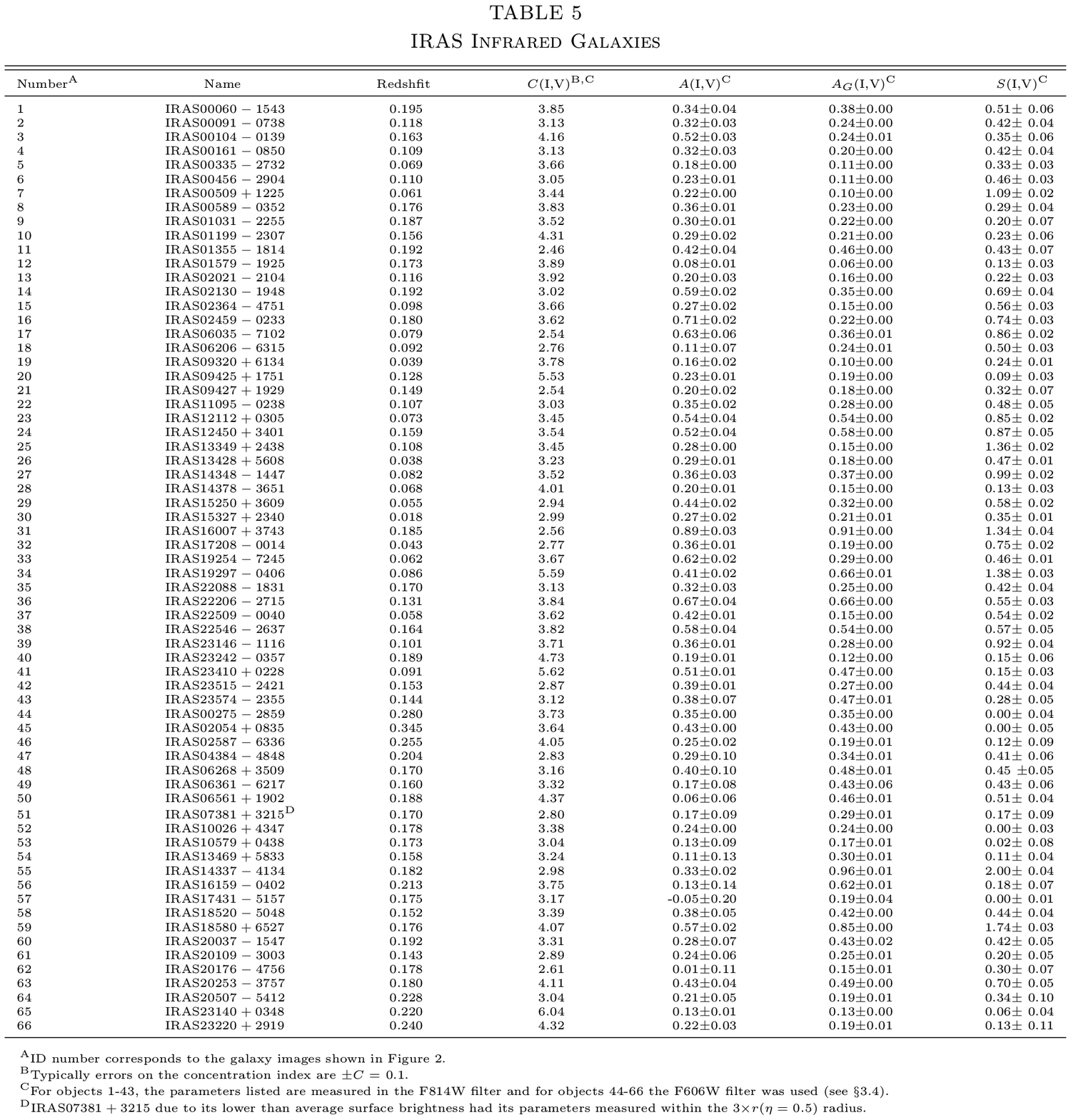}{6.0in}{0}{100}{100}{-310}{-170}
\end{figure}

\begin{figure}
\plotfiddle{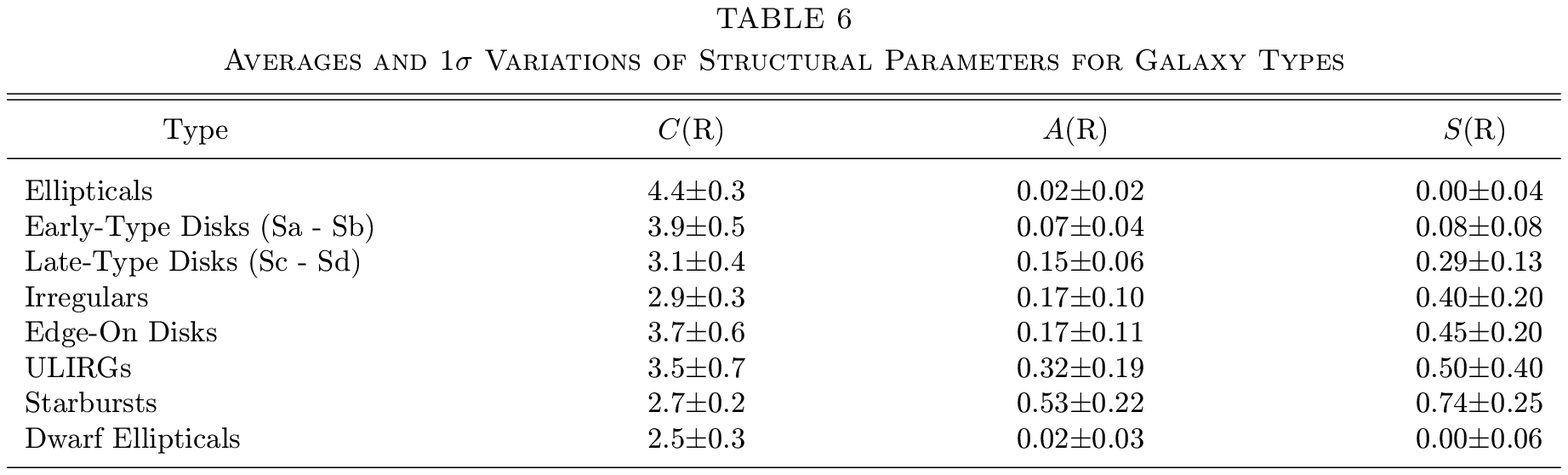}{6.0in}{0}{100}{100}{-310}{-170}
\end{figure}

\begin{figure}
\plotfiddle{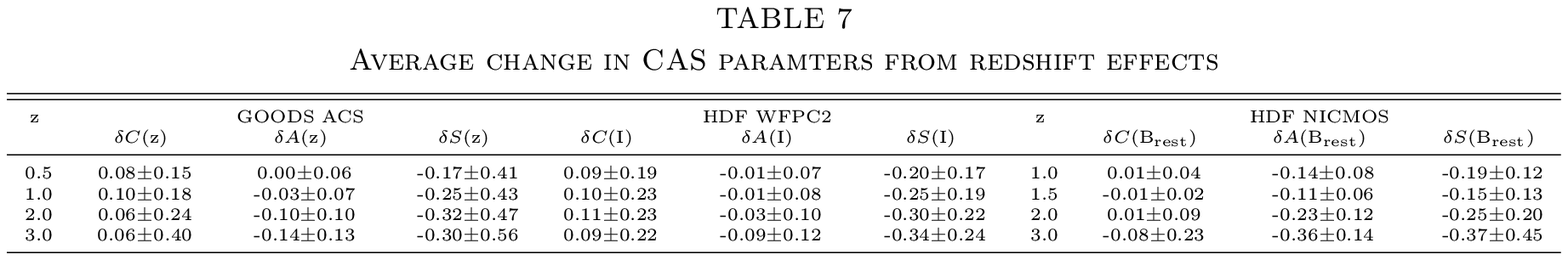}{6.0in}{0}{100}{100}{-310}{-170}
\end{figure}

\end{document}